%% file: forgemegakernel.tex
\documentclass[sigplan,nonacm]{acmart}

\renewcommand\footnotetextcopyrightpermission[1]{}
\makeatletter
\AtBeginDocument{%
  \fancypagestyle{standardpagestyle}{%
    \fancyhf{}%
    \renewcommand{\headrulewidth}{\z@}%
    \renewcommand{\footrulewidth}{\z@}%
    \fancyfoot[C]{\if@ACM@printfolios\footnotesize\thepage\fi}%
    \if@ACM@review
      \fancyhead[LO,LE]{\ACM@linecountL}%
      \fancyhead[RO,RE]{\ACM@linecountR}%
    \fi
  }%
  \pagestyle{standardpagestyle}%
}
\makeatother
\usepackage{booktabs}
\usepackage{array}
\usepackage{float}
\floatstyle{ruled}
\newfloat{algorithm}{tbp}{loa}
\floatname{algorithm}{Algorithm}
\usepackage{multirow}
\usepackage{graphicx}
\usepackage{amsmath}
\usepackage{enumitem}
\usepackage{xcolor}
\usepackage{tikz}
\usetikzlibrary{arrows.meta,positioning,fit,backgrounds,calc,decorations.pathreplacing}

\newcommand{\sys}{\textsc{ForgeMegakernel}}
\newcommand{\eqcite}[1]{Eq.\,\eqref{#1}}
\newif\ifdraftcolours \draftcoloursfalse

\newcommand{\draftname}[1]{\ifdraftcolours\textcolor{cSglang}{#1}\else#1\fi}
\newcommand{\rev}[1]{\ifdraftcolours\textbf{\textcolor{cOurs}{#1}}\else#1\fi}

\newcommand{\failB}[1]{\ifdraftcolours\textcolor{cHazy}{#1}\else#1\fi}
\newcommand{\failC}[1]{\ifdraftcolours\textcolor{cMpk}{#1}\else#1\fi}
\newcommand{\rung}[1]{\textsf{#1}}
\newcommand{\code}[1]{{\texttt{\small\hyphenpenalty=0\exhyphenpenalty=0\relax #1}}}
\newcommand{\us}{\,$\mu$s}

\definecolor{cOurs}{HTML}{CC3311}     
\definecolor{cOurs2}{HTML}{EE99AA}    
\definecolor{cOurs3}{HTML}{661000}    
\definecolor{cSglang}{HTML}{0077BB}   
\definecolor{cMpk}{HTML}{9B2F8F}      
\definecolor{cHazy}{HTML}{009988}     
\definecolor{cAccent}{HTML}{33BBEE}
\definecolor{cGrey}{HTML}{4D4D4D}
\definecolor{cMute}{HTML}{A2A8B0}
\definecolor{cInk}{HTML}{1B2128}

\setlist[itemize]{leftmargin=1.1em,topsep=2pt,itemsep=1pt,parsep=0pt}
\setlist[enumerate]{leftmargin=1.4em,topsep=2pt,itemsep=1pt,parsep=0pt}

\begin{document}

\title{\sys: A General Framework for Efficient Auto-Regressive Model Decode Megakernels}

\author{Leshan Li}
\affiliation{\institution{Tsinghua University}\city{Beijing}\country{China}}
\email{lils25@mails.tsinghua.edu.cn}

\author{Zhui Zhu}
\affiliation{\institution{Tsinghua University}\city{Beijing}\country{China}}
\email{z-zhu22@mails.tsinghua.edu.cn}

\author{Xianglong Deng}
\affiliation{\institution{University of Chinese Academy of Sciences}\city{Beijing}\country{China}}
\email{dengxianglong02@gmail.com}

\author{Yaojian Chen}
\affiliation{\institution{ModelBest Inc.}\city{Beijing}\country{China}}
\email{chenyaojian17@gmail.com}

\author{Qingfeng He}
\affiliation{\institution{Tsinghua University}\city{Beijing}\country{China}}
\email{heqf21@mails.tsinghua.edu.cn}

\author{Yuxuan Li}
\authornote{Corresponding authors.}
\affiliation{\institution{ModelBest Inc.}\city{Beijing}\country{China}}
\email{liyuxuan@modelbest.cn}

\author{Rong Zhao}
\affiliation{\institution{Tsinghua University}\city{Beijing}\country{China}}
\email{r\_zhao@tsinghua.edu.cn}

\author{Xu Han}
\authornotemark[1]
\affiliation{\institution{Tsinghua University}\city{Beijing}\country{China}}
\email{han-xu@mail.tsinghua.edu.cn}

\author{Zhiyuan Liu}
\authornotemark[1]
\affiliation{\institution{Tsinghua University}\city{Beijing}\country{China}}
\email{liuzy@tsinghua.edu.cn}

\begin{abstract}
\input{sec/00-abstract}
\end{abstract}


\maketitle

\input{sec/01-intro}
\input{sec/02-background}
\input{sec/03-observations}
\input{sec/04-design}
\input{sec/05-evaluation}

\input{sec/06-discussion}
\input{sec/07-related}
\input{sec/08-conclusion}

\bibliographystyle{ACM-Reference-Format-cite}
\bibliography{refs}

\appendix

\section{Milestone ablation}
\label{sec:eval:abl}

\input{sec/abl-body}

\section{A second coding agent}
\label{sec:eval:supp}

\input{sec/supp-body}

\end{document}

%% file: sec/00-abstract.tex
Auto-regressive model decode is bandwidth-bound, since every weight and
key/value-cache byte crosses high-bandwidth memory once per token. A megakernel
is an ideal solution, but existing automatic megakernel generation approaches cannot achieve both
generalization across models and correctness guarantees.

We present \sys{}, which generates a per-model high-performance decode
megakernel using coding agents. \sys{} pairs a universal knowledge base of ten progressive milestones with
an independent mid-state test oracle. The milestones provide the megakernel's
structural properties: a fine-grained instruction stream for each SM,
dependency counters replacing the global synchronization, and a shared-memory
buffer pool for workload balance across SMs and greater parallelism. The test oracle derives the mid-states of the megakernel and checks the
performance, error and precision during the generation process, guaranteeing a
correct and trustworthy forged megakernel.

We evaluated \sys{} on 14 representative decoding operations across eight
model families spanning 0.6B--13B parameters. The generated megakernels
achieved 50.5--85.9\% MBU and geometric mean speedups of $1.21\times$ over
SGLang~0.5.18 and $1.54\times$ over a megakernel compiler under identical
configurations. Inside SGLang, evaluated on GSM8K with ragged prompts, all 14
megakernels decoded faster than the SGLang engine at comparable answer
accuracy.

%% file: sec/01-intro.tex
\section{Introduction}
\label{sec:intro}

A megakernel executes an entire computation in a single persistent kernel launch instead of one launch per operator~\cite{gupta2012persistent,laine2013megakernels}. The most important application is auto-regressive model decoding, which dominates the latency of interactive services. Decoding generates one token at a time by reading model weights and performing matrix-vector multiplications~\cite{yu2022orca,agrawal2024sarathi}, an arithmetic intensity near one floating-point operation per byte that leaves GPU arithmetic units idle and makes decoding memory-bound.


Current serving engines execute each decoding step as a sequence of
operator-specific kernels. For example, vLLM~\cite{kwon2023pagedattention} and
SGLang~\cite{zheng2024sglang} construct each step using precompiled library
kernels~\cite{nvidia2025cublas,nvidia2025cutlass,tillet2019triton,ye2025flashinfer},
launching one kernel per operator. On an H100 GPU, this approach requires
280--448 kernel launches per decoded token. \rev{Each kernel launch incurs startup and retirement overhead, making the
decoding step take 1.06--1.45$\times$ than a single persistent-kernel
execution that transfers the same number of bytes.  Moreover, in the 280-launch case, 158 launches
execute small RMSNorm, RoPE, or KV-store kernels. These kernels induce memory transfer of intermediate values between HBM and shared memory, which accounts for one-fifth of the per-token decoding latency.  }
(\S\ref{sec:eval:profile})


A decode megakernel removes this overhead by fusing the entire decoding step into that single launch and executing the small operations in place within the thread blocks. Handwritten implementations~\cite{spector2025nobubbles,nrusimha2025flashformer}, built on optimized operators and tile-based programming languages~\cite{dao2022flashattention,dao2024flashattention2,spector2025thunderkittens}, achieve 74--78\% of H100 memory bandwidth. \failB{However, these implementations lack reusability: each targets a single model, and adapting it to a different model requires weeks of specialist effort}.

For rapid development, several works~\cite{mpk2025,eventtensor2026,adamk2026,ouyang2025kernelbench,astra2025,cudaforge2025,kernelforge2026} focus on automatic megakernel generation. These works fall into two categories, namely compiler-based~\cite{mpk2025,eventtensor2026,adamk2026} and agent-driven~\cite{ouyang2025kernelbench,astra2025,cudaforge2025,kernelforge2026}. However, \failC{neither consistently produces high-performance megakernels on
arbitrary cells (model size, batch size, sequence
length)}.\footnote{A \emph{cell} is one (model, batch size, sequence length)
triple, the configuration a decode megakernel is compiled and measured for.} Compiler-based
approches~\cite{mpk2025,eventtensor2026,adamk2026} follow strict lowering rules
and exhibits performance degradation on the cells those rules do not cover. Agent-driven approaches
generate a megakernel for any cell in this
space~\cite{ouyang2025kernelbench,astra2025,cudaforge2025,kernelforge2026}.
However, with no guide or knowledge for the generation task, the generated kernel is
generally slower than the serving engine (\S\ref{sec:obs:cell}). 
We view agent-driven generation as a promising approach for arbitrary cells; the missing component is a knowledge base that guides high-performance megakernel generation across auto-regressive models. This leads to \emph{Challenge~1: how to formulate a knowledge base that supports agent generalization, contains no model-specific parameters, and is precise enough to guide implementation and verification.}

Guaranteeing the correctness of an agent-generated megakernel remains an
open problem. A megakernel fuses the operators of a decoding step,
including RMSNorm, attention, and projection operations, keeping
intermediate results within the kernel and preventing direct external
inspection. Checking only the final output leaves the fusion process
unchecked and prone to errors, while output discrepancies do not identify
where those errors originate. Moreover, if a correctness check is
specified in the prompt but implemented and executed by the agent, the
build is gated by an agent-defined check rather than a standardized one.
Loopholes in this check may allow the agent to bypass the intended
correctness requirements (\S\ref{sec:obs:oracle}).
This leads to \emph{Challenge~2: recovering verifiable mid-states
of the megakernel and validating them through a standardized
correctness gate during megakernel generation.}

We present \sys{}, an end-to-end framework that generates a deployable
decoding megakernel from a model configuration. Generation follows a
sequence of ten milestones, each specifying a structural property of a
high-performance decoding megakernel and a diagnostic for evaluating
that property. These milestones contain no model-specific parameters
and thus apply across all cells in the target space (Challenge~1).
A mid-state test oracle serves as the validation gate, deriving
all quantities and contracts required for evaluation, including the
byte counts underlying MBU, precision contracts, and a float64 error
bound. The oracle performs these checks during megakernel generation,
evaluating every kernel against a uniform standard rather than
agent-defined checks (Challenge~2). An iterative loop coordinates
milestone-guided generation and oracle-based validation, with GPU
execution in every iteration.

In each iteration, a newly initialized coding
agent receives the milestone matrix, the previous gate summary, and the
campaign rules. The coding agent modifies the kernel, runs the validation
gate on an H100 GPU, and retains or reverts the changes based on the
measured results. Only measured values are carried forward to the next
round; the hypothesis, code diff, and measurement results are recorded
in a ledger. Before the next milestone becomes available, a second agent
audits each round that passes the gate, using the code diff and run
identifiers as evidence. 

In this work, we make the following contributions:

\begin{enumerate}[label=\textbf{C\arabic*.},leftmargin=2.6em]
\item \textbf{We propose a universal knowledge base comprising ten milestones that specify the structural properties of high-performance megakernels for auto-regressive models.} Each ordered milestone identifies a structural property, including a single persistent launch, a per-SM instruction stream, counter dependencies without a grid-wide barrier, and shared-memory buffer pooling (\rung{M5}--\rung{M8}). Each milestone also specifies an MBU floor. A milestone constrains the kernel structure without fixing parameter values. For a given model, the agent derives all required constants, allowing the same ten milestones to apply across the entire design space (\S\ref{sec:design:ladder},~\S\ref{sec:eval:kernel}).

\item \textbf{We propose a standardized test oracle for validating megakernel mid-states during construction, thereby ensuring correctness and trustworthiness.} The oracle reads the mid-state key/value cache, enabling divergence to be localized within the kernel. The oracle also derives every quantity used in evaluation from the model configuration. This design prevents the agent from altering the performance threshold against which the kernel is assessed. Accuracy inside SGLang matches that of SGLang's own decoder for every megakernel included in our evaluation (\S\ref{sec:design:oracle},~\S\ref{sec:eval:acc}).

\item \textbf{We develop an end-to-end generation framework that incorporates feedback from the GPU.} Each iteration runs on the device. A newly initialized agent edits the kernel, the evaluation gate returns measured latency, utilization, profiler counters, and error information, and the subsequent edit is guided by these measurements rather than by estimates. An insertion path registers the completed kernel in a production serving engine without further modification (\S\ref{sec:design:loop},~\S\ref{sec:eval:insertion}).
\end{enumerate}

We evaluate \sys{} at the kernel and serving levels. The megakernel
generation campaigns produced 14 megakernels for eight models (0.6B--13B
parameters), with MBU ranging from 50.5\% to 85.9\%. On byte-identical
measurement cells, the megakernels achieved geometric-mean speedups of
$1.21\times$ over SGLang~0.5.18 and $1.54\times$ over a megakernel
compiler.

When integrated into SGLang, all 14 megakernels reduced end-to-end
latency. When evaluated against an independent float64 reference, all
megakernels exhibit comparable accuracy compared with SGLang. In serving-level evaluations, all megakernels
 remain consistent with SGLang in GSM8K accuracy.

In the ablation study, removing a milestone's structural requirement while
retaining its MBU floor reduced the final campaign result for every pair tested
(\S\ref{sec:eval:abl}).

%% file: sec/02-background.tex
\section{Background}
\label{sec:bg}

\subsection{The decode step}
\label{sec:bg:decode}

Auto-regressive model serving splits into prefill, which processes the prompt in parallel and is
compute-bound, and decode, which generates one token at a time and is
bandwidth-bound at interactive batch
sizes~\cite{yu2022orca,agrawal2024sarathi,kamath2025podattention}.

One decode step of a
auto-regressive transformer~\cite{vaswani2017attention} applies per layer RMSNorm, the
$QKV$ projections, RoPE~\cite{su2024rope}, attention, the $O$ projection and a
gate/up--SiLU--down MLP, then a final RMSNorm and the LM head. Every projection
reads a whole weight matrix; attention alone reads state from earlier steps, the
whole key/value (KV) cache per token, with $q_j$ the query of head $j$ and
$K_{g(j)},V_{g(j)}\in\mathbb{R}^{S\times d_h}$ the cache of its KV head $g(j)$
over $S$ positions:
\begin{equation}
\label{eq:attn}
a_j = \mathrm{softmax}\!\left(\frac{q_j K_{g(j)}^{\top}}{\sqrt{d_h}}\right)
V_{g(j)},
\end{equation}

\subsection{Model bandwidth utilization}
\label{sec:bg:bandwidth}

For a model with $L$ layers, hidden size $h$, $n_h$ query heads, $n_{kv}$
grouped key/value heads~\cite{shazeer2019mqa,ainslie2023gqa}, intermediate width
$i$ and vocabulary size $V$, summing the operators that read memory, one decode
step at batch size $b$ reads

\begin{equation}
\label{eq:bytes}
\begin{split}
B(b,S) = \underbrace{\left[L(h n_h d_h + 2 h n_{kv} d_h + n_h d_h h + 3 h i) + V h\right] w}_{\text{weights}} \\
{} + \underbrace{2 L\,n_{kv} d_h S b\,w}_{\text{KV cache}} + \underbrace{b\,h\,w}_{\text{embedding rows}}
\end{split}
\end{equation}

\noindent bytes from high-bandwidth memory (HBM), where $w$ is the element
width: the projections and the LM head, the attention reads of the cache, and
the gathered embedding rows. The embedding
table is excluded from \eqcite{eq:bytes} because decode gathers $b$ rows rather
than reading the full table. Including the full table can inflate reported
utilization by up to $1.22\times$ for untied embeddings. We observed this
over-count in the reported measurements of two published megakernel systems.

Model bandwidth utilization (MBU)~\cite{databricks2023mbu} is defined as
\begin{equation}
\label{eq:mbu}
\mathrm{MBU} = \frac{B(b,S)}{t \cdot \beta_{\text{peak}}},
\end{equation}
where $t$ is the wall-clock time per generated token and $\beta_{\text{peak}}$
is peak HBM bandwidth. We use an H100 80GB with HBM3 and
$\beta_{\text{peak}}=3350$\,GB/s throughout. Because bandwidth is the binding
resource during decode, the 100\% reference applies to bandwidth rather than to
an unsaturated resource. At fixed $(b,S)$, $B(b,S)$
is determined by the model and measurement configuration, so an MBU ratio is also a
latency ratio.

\begin{figure*}[!t]
\centering
\includegraphics[width=\textwidth]{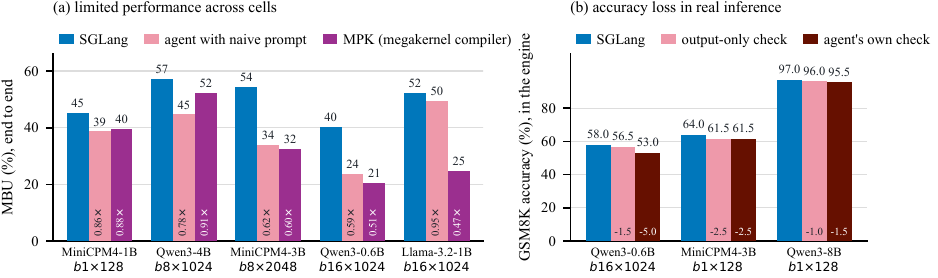}
\caption{One H100, identical bytes. \textbf{(a)} five cells under SGLang, an
agent given a naive prompt, and a megakernel compiler (MPK), ratios against
SGLang. \textbf{(b)} three cells scored on GSM8K against SGLang. The
\emph{output-only check} reads the final logits alone, the \emph{agent's own
check} is one session told in its prompt to guarantee correctness, with no
harness and no gate.}
\label{fig:obschain}
\end{figure*}

\subsection{Megakernels}
\label{sec:bg:mega}

A conventional serving engine executes a decode step as several hundred kernels,
launched individually or replayed from a captured CUDA graph as vLLM and SGLang
do by default~\cite{nvidia2024cudagraphs}. Graph replay removes fixed launch
overhead but not the transition between kernels: the memory system must finish
one grid before the next can make progress. A megakernel keeps the step within
one host launch, so the weight stream never stops at an operator boundary and
the weightless operators of \S\ref{sec:bg:decode} run in place.

Published megakernels are point
designs~\cite{spector2025nobubbles,nrusimha2025flashformer,mpk2025}. Each holds
the GPU's streaming multiprocessors (SMs) for the whole step and arranges the
work among them inside the kernel, but how it does so, and the constants that go
with it, are settled for one model in advance: a hand-written megakernel reaches
74--78\% of H100 bandwidth on the one model it was written
for~\cite{spector2025nobubbles}, and the compiler-based design lowers any model
through rules fixed before it. Neither
says what a kernel for a model it has not seen must look like.

Persistent threads originated in general-purpose GPU
computing~\cite{gupta2012persistent}, and graphics systems documented their
drawbacks for monolithic kernels~\cite{laine2013megakernels}; decode's
regularity, the same matrix--vector products and normalizations applied to a
weight stream, is what makes the approach apply here.


%% file: sec/03-observations.tex
\section{\draftname{Observations}}
\label{sec:obs}

We evaluate the performance of megakernels produced by two automatic generation
approaches on five cells (\S\ref{sec:obs:cell}), and examine what a correctness
check placed at the model output accepts (\S\ref{sec:obs:oracle}). Every run uses
one H100 and the byte count of
\eqcite{eq:bytes}; Figure~\ref{fig:obschain} reports both measurements.

\subsection{Limited performance across cells}
\label{sec:obs:cell}

Two approaches generate a decode megakernel without a specialist: a coding agent
asked for one directly, and a compiler that lowers the model through fixed rules.
We measure both against SGLang on five cells, at one byte count and card.

\textbf{Agent under a naive prompt.} An agent given no structural target writes a kernel that runs but does not run
fast. Our naive prompt names the model, the cell and the single-launch
requirement, and no structural property of the kernel; the agent produced a
working megakernel on all five cells, and all five are slower than SGLang: 0.59
to 0.95 times its utilization, geometric mean 0.75 (Fig.~\ref{fig:obschain}a). They lose their time at different places: one leaves 28\% of the
machine idle inside the kernel; one retains 225 grid-wide barriers in a step,
synchronized through a single dependency counter; one spends more time selecting
a token than computing the logits. A bandwidth target is no substitute, since MBU
is one figure over a fused step and does not indicate which part to change: in
one campaign the agent kept a grid-barrier phase structure for 35 measured rounds
and stopped at 34.2\%, above the 31\% SGLang reaches on that cell and well below
a hand-written megakernel's 74.5\% on the same card.

\textbf{Compiler under fixed rules.} A compiler is repeatable, but its rules cover some cells better than others. MPK
lowers the same five cells and is no faster than the agent, at 0.47 to 0.91 times
SGLang's utilization with a geometric mean of 0.65 (Fig.~\ref{fig:obschain}a),
and the lowering is brittle where a shape crosses a rule: one line in a hand-written task kernel pads the
batch to 8 or to 16, and MPK's utilization falls by about 44\% when a ninth
sequence enters a batch of eight.

The above experiments show that fixed compilation rules have limited coverage
across cells, while an agent without structural guidance cannot reliably find
effective optimization directions; an aggregate bandwidth metric does not
indicate which structure should be improved. High-performance megakernel
generation therefore requires a set of reusable, structured milestones that
provide the agent with explicit implementation goals and optimization steps.

\subsection{Unverifiable correctness of kernel fusing}
\label{sec:obs:oracle}

A generated megakernel is checked in one of two ways today: its final output is
compared against a reference, or the check is stated in the prompt and the agent
establishes correctness itself. Fusing keeps every internal value in registers and
shared memory, so the logits are the only surface an outside checker can read.

\textbf{The output-only check.} An output-only check cannot localize the
defects it does detect. A kernel
that broadcasts one sequence's decode position across a ragged batch is
indistinguishable from a correct one whenever the prompts have equal length, and
a single distance at the logits cannot identify which of the $7L$ fused operator
sites of an $L$-layer model produced an error. Kernels accepted under such a
check are measurably worse in the serving engine: the \emph{output-only check}
kernels of Fig.~\ref{fig:obschain}b lose 1.5 GSM8K points on Qwen3-0.6B at
$b16{\times}1024$, 2.5 points on MiniCPM4-3B at $b1{\times}128$ and 1.0 point on
Qwen3-8B at $b1{\times}128$, each against the SGLang engine on the same
prompts.

\textbf{The agent's own check.} Stating the check in the prompt does not recover
the missing mid-state signals. The
\emph{agent's own check} kernels of Fig.~\ref{fig:obschain}b come from single
sessions asked to guarantee the correctness of the megakernel they wrote, with
no harness and no gate. Each session wrote its own tests and cleared an MBU bar
set at SGLang's utilization on its cell, and each kernel still lost accuracy
against the SGLang engine: 5.0 GSM8K points on Qwen3-0.6B at $b16{\times}1024$,
2.5 points on MiniCPM4-3B at $b1{\times}128$ and 1.5 points on Qwen3-8B at
$b1{\times}128$. Besides, no two agents write the same check
gate, so an agent's own check cannot be standardized and is
therefore hard to trust.

The above experiments show that final-output checks may miss specific defects
in a fused kernel and cannot identify their sources, while acceptance criteria
defined by the generating agent may treat defects as acceptable limitations.
Correctness validation therefore requires observable intermediate states and
criteria independent of the generating agent to detect and localize errors in
the internal computation.

%% file: sec/04-design.tex
\section{Design}
\label{sec:design}

\input{sec/fig-arch}

\sys{} takes a model configuration and returns a decode megakernel a serving
engine can load. \S\ref{sec:obs} leaves two obstacles: an agent has no statement
of what a fast megakernel looks like, and a fused kernel offers nothing to check
against. The milestone ladder (\S\ref{sec:design:ladder}) supplies the first and
the mid-state test Oracle (\S\ref{sec:design:oracle}) the second, deriving every
quantity a verdict rests on before a kernel exists; the generation framework
(\S\ref{sec:design:loop}) drives both and deploys the forged megakernel in a
serving engine (Figure~\ref{fig:arch}).

\subsection{The universal knowledge base and the milestones}
\label{sec:design:ladder}

Building a decode megakernel takes two kinds of progress, preserving the model's
computation as operators fuse and reorganizing execution so that fusion keeps
the GPU busy, and \sys{} takes the two in order, first establishing correctness
from individual operators through the full decoder, then bringing the entire
step, including the LM head, into one persistent cooperative launch kernel, a
megakernel. The single launch removes per-operator launch boundaries. However,
data dependencies between SMs remain, so the megakernel can still run one
operator at a time behind grid-wide barriers, and \S\ref{sec:obs:cell} shows
that an agent tuning inside that barrier-serialized structure, with no knowledge
or guide, stays well below the roofline. We divide the tuning work into three
steps: what each SM executes, when that SM may proceed, and where the SM's
operands are staged.

The first step is a per-SM instruction stream. The host compiles the decoder
into typed instructions, each carrying explicit operand descriptors, and assigns
one stream to each SM; an interpreter inside the kernel dispatches those
instructions. SMs then advance through different work instead of sharing a
single program position. This step is a precondition rather than a gain: the
instruction stream grants the scheduling freedom the next two steps require, but
by itself removes no wait and hides no memory latency.

The second step is a dependency counter. Separate streams help only when an SM
waits for the instructions that produce its inputs rather than for the whole
grid. A counter semaphore records each such edge: a producer increments the
counter on completion, and a consumer proceeds once its own inputs are ready. Replacing grid-wide barriers with these counters converts the
scheduling freedom of the first step into independent progress. Neither step
suffices alone. Without counters, the slowest SM still sets every phase
transition; with counters over a coarse phase schedule, the same waits return
under another name.

The third step is a shared-memory buffer pool. Independent progress still exposes
memory latency while each instruction loads, computes and stores in turn. A buffer
pool provides staging space whose lifetime follows the instructions that use
that space, so the loads of instruction $k{+}1$ overlap the stores of
instruction $k$ and no buffer is reclaimed before its consumers finish. Without
the pool, independent streams need not form a memory pipeline; buffer pooling therefore
follows scheduling and dependency management rather than standing on its own.

Together the three steps close the decomposition: instruction streams assign
work, counters order that work, and buffers hold the operands whose movement
consecutive instructions overlap, each step covering an obligation the other two
leave open. This sufficiency is structural. We do not claim that these
implementations are the only ones, nor that satisfying the three obligations
guarantees a given bandwidth, since scheduling balance, instruction granularity
and prefetch distance still have to be tuned. The ladder therefore requires
structure and measured performance together.

These dependencies become the ten milestones of Table~\ref{tab:ladder}. The
functional prefix validates the toolchain, individual operators, a fused layer,
the $L$-layer stack and decode on a real checkpoint against the float64 golden.
The next milestone establishes one persistent launch, checking the launch count
and forbidden library calls so that no operator, the LM head included, remains
outside. The three steps above form the performance core, each adding a
structural requirement and raising the MBU floor. The final milestone is an
optional stability run: 1000 greedy tokens with no NaN or infinity, at most five
positions of drift from the bf16 reference, and a free run to 4000 tokens. The
stability run targets index overflow, buffer-pool fragmentation and counter
wraparound.

Structural gates inspect the source and the host schedule, and measure each
structural effect that is observable. The interpreter gate requires at least
five instruction types and six instructions per layer, with queue imbalance
within $1.35\times$ the mean, which rules out an interpreter that assigns most
work to one SM. The dependency gate requires zero grid-wide barriers, at least
$4L$ distinct counters, $\text{busy\_frac} \ge 0.85$ (instruction spans over
$\text{SMs} \times t$) and $\text{tail\_spread} \le 0.05$ (first-to-last SM
finish), both read from a \code{\%globaltimer} timeline the gate holds within
25\% of the step's CUDA-event latency. The buffer-pooling gate requires shared-memory
buffer allocation. Algorithm~\ref{alg:gate} states these checks alongside the
performance tests of the same milestone.

\input{sec/tab-ladder}

A campaign always works on its lowest unpassed milestone. A fallback pass
records partial progress, but only the target threshold opens the next
milestone. Every gate returns the same envelope, \code{\{status, suite, summary,
metrics, details\}}, so each round receives both the obligation the campaign has
not met and the measurements to act on.

\label{sec:obs:constants}
The ladder specifies execution properties, not a kernel template or a language
above CUDA. Layer count, head count and batch size change the instruction
streams, the dependency graph and the storage sizes, but not the three
obligations; the agent derives those quantities from the model dimensions and
writes the kernel from scratch. One
ladder therefore guides every auto-regressive-model campaign, while each gate checks the
implementation for the model at hand.

\subsection{The mid-state test oracle}
\label{sec:design:oracle}

As discussed in Section~\ref{sec:obs:oracle}, kernel fusion removes the operator
boundaries at which an implementation is normally compared against a reference,
leaving only the logits visible to an external checker. This makes the existing
verification methods insufficient: such methods cannot inspect the intermediates
where many errors originate. We therefore propose a mid-state test oracle
that compares a kernel's intermediates with an external reference, placing the
acceptance criteria beyond the agent's reach. The oracle has three complementary
measurements, alongside a set of tolerance-free functional checks
(Figure~\ref{fig:arch}): (i) a byte-accounted performance measurement, which checks whether a
kernel appears fast only because it moves fewer bytes than the model requires;
(ii) a float64-referenced error measurement, which checks whether a kernel
appears close to its reference only because the reference rounds wherever the
kernel does; and (iii) a precision contract (a field consensus that certain
operators, such as softmax and the accumulators, must not be narrowed below
fp32) on arithmetic width, which checks whether a kernel produces the right
output while computing at the wrong width (Algorithm~\ref{alg:gate}).

All three oracle components use configuration-derived quantities that the host
fixes before kernel generation. The host combines the task configuration
(architecture, dimensions, dtype, and gated $(b,S)$ cells) with the machine
configuration (GPU, transport, and peak HBM bandwidth). From the combined
configuration, the host computes the parameter count, the byte count $B(b,S)$
defined in \eqcite{eq:bytes}, the KV share, and all tolerances specified below,
then transmits these quantities with hashes. Each gate receives only the required
quantities and is prohibited from recomputing those quantities. To protect the
evaluator, the agent's write access is restricted to the kernel source and notes;
any round that modifies the evaluator is reverted.

In addition to the configuration-derived tolerances, the protected evaluator
enforces three functional checks from \rung{M5} onward. The three functional
checks admit no tolerance: numerical agreement with a reference does not substitute
for passing the functional checks. First, the logit-finiteness check requires
all logits to be finite.
Second, the KV-read-boundary check requires poisoning the KV arena past the
write cursor with $\sigma=100$ noise to leave the logits bit-identical, since
an out-of-bounds read may otherwise produce plausible outputs. Third, the
per-lane-position check requires lanes at different positions to decode at
the position assigned to each lane, rather than all using lane~0's position.

Besides the functional checks, the oracle evaluates efficiency using
byte-accounted performance measurements. Since self-reported utilization
depends on a backend's own accounting, the oracle accepts only elapsed
times obtained through host-assisted measurement.
The host calculates the byte count $B(b,S)$ required per
decoding step from the model configuration, which includes activated
parameters, accessed cache entries, and gathered embedding rows.
The per-step latency is estimated as the two-point slope
$t = \frac{\min_r\{t_r(80) - t_r(16)\}}{64}$ across replicates $r$,
eliminating prefill and all length-independent costs.
MBU is then computed using
\eqcite{eq:mbu} and the host-derived $B(b,S)$.
To ensure measurement correctness, three additional independent checks
are performed: (i) Nsight Compute counters measure
the bytes a kernel actually moves, from which we define the read
amplification $\rho$ as the ratio of those measured bytes to $B(b,S)$.
Moving $\rho B(b,S)$ bytes takes at least
$\rho B(b,S)/\beta_{\text{peak}}$ seconds under the
bandwidth roofline~\cite{williams2009roofline},
so \eqcite{eq:mbu} bounds MBU by $\rho^{-1}$, giving an upper bound
of approximately $71.4\%$ for $\rho = 1.4$ regardless of execution speed.
This check measures redundant traffic directly
rather than inferring it from latency; (ii) the gate enforces
$\text{MBU} \le 100\%$ to detect byte counts that include data
the step never accesses; counting the full embedding table alone
violates this bound by $1.22\times$;
(iii) each backend declares its timing scope as
\texttt{kernel-only}, \texttt{step-no-sampling}, or \texttt{end-to-end},
ensuring that comparisons use the same scope.

\input{sec/fig-gate}

Performance measurements establish how fast a kernel runs, but not how
accurately it computes. The oracle therefore checks numerical error in both
the intermediate KV cache and the final logits.
Both error surfaces are read against a float64 forward pass, exactly upcast from
the canonical bf16 checkpoint with nothing rounded in between. A fault must be located before it can
be fixed, and the only intermediate that must reach global memory is the KV
cache. The harness owns that buffer, so the gate reads the cache per layer without
instrumenting the agent's code, and the rows arrive in execution order, which
locates the first divergence. The rows are compared against a PyTorch implementation
in the kernel's own arithmetic, since a bf16 tensor differs from exact arithmetic
by about $10^{-2}$ whatever the kernel does, a floor that hides real defects;
float64 sets the scale of the comparison instead. Over the $2L$ rows
$\mathcal{R}$, one key and one value per layer, with $|r|$ elements in row $r$
and $\epsilon = 10^{-6}$ in the denominator, write
\begin{equation}
\label{eq:cb}
\delta(X,Y) = \frac{1}{|\mathcal{R}|}\sum_{r\in\mathcal{R}} \frac{1}{|r|}
  \sum_{j\in r}
  \frac{2\,\lvert X_{rj} - Y_{rj}\rvert}
       {\lvert X_{rj}\rvert + \lvert Y_{rj}\rvert + \epsilon}.
\end{equation}
With $K$ the kernel, $P$ the PyTorch reference and $G$ the golden, the bar is
$C < 2B$ for $C = \delta(K,P)$ and $B = \delta(P,G)$: the kernel's distance from
the reference implementation may not exceed twice that implementation's distance
from exact arithmetic.

While the KV-cache comparison locates intermediate divergences, the logit-level
check measures their effect on the output distribution.
At the logits the golden is used directly, on the
log-softmax over the golden's own top 64 entries, at teacher-forced taps along that continuation, which keeps the taps independent, since a kernel that diverges once
is answering a different prompt from that point on. With $\mathcal{T}$ the taps,
$\mathcal{K}_t$ the golden's top 64 entries at tap $t$ and
$\ell = \log\mathrm{softmax}$, the distance of an engine $E$ from the golden is
\begin{equation}
\label{eq:eps}
\varepsilon(E) = \frac{1}{|\mathcal{T}|}\sum_{t\in\mathcal{T}}\;
  \frac{1}{64}\sum_{v\in\mathcal{K}_t}
  \bigl\lvert \ell(E_t)_v - \ell(G_t)_v \bigr\rvert ,
\end{equation}
and $\varepsilon^{\max}$ is the same quantity taking maxima in place of both
means. Writing $D = \varepsilon(\text{HF bf16})$ for the bf16 reference's own
distance on this surface, measured in the campaign's own oracle build, the three
bars are $\varepsilon(K) \le 5D$,
$\varepsilon(K) \le \varepsilon(\text{SGLang})$ and
$\varepsilon^{\max}(K) \le 2\,\varepsilon^{\max}(\text{SGLang})$. Each is either
what bf16 costs on this model or what the production engine achieves on the same
taps, so none is a value we chose. Greedy agreement is scored against a floor
derived the same way, and where no float64 family exists the check abstains
rather than passes.

An error bound scores the output and says nothing about the arithmetic behind that output,
and the two can move in opposite directions, since rounding an intermediate to
bf16 moves a kernel toward any reference that rounds there too. The operator
tolerance leaves room for that move, at $2\times10^{-2}$ against a bf16 ulp of
$1.95\times10^{-3}$. The oracle therefore stores a precision contract read-only, bf16 at
operator boundaries only and fp32 between an operator's input and its output,
including accumulators, norm gains and the rotary table. The gate tests the
contract per boundary-carrying operator. From one operator's tap inputs it
builds two references, $A$ computed to the contract and $A'$ computed with that
operator's intermediate narrowed to bf16, and writes their difference
$d = A' - A$, the rounding under test. With $K$ the kernel's output on those
same inputs, the gate regresses $K - A$ onto $d$,
\begin{equation}
\label{eq:alpha}
\alpha = \frac{\langle K - A,\; d\rangle}{\langle d,\; d\rangle},
\qquad \text{pass if } \alpha \le 0.5 .
\end{equation}
Here $\alpha$ is the fraction of that extra rounding the kernel adopted, $0$ for
the contract and $1$ for the narrowed variant. The unit of $\alpha$ is the deviation being tested for, so the threshold needs no calibration, and reduction reordering moves
the output in an unrelated direction without raising $\alpha$. The check abstains
if the two references coincide.

Taken together, the three parts leave the agent nothing to redefine: the byte
count, the float64 references and the arithmetic widths are all fixed by the
model configuration before a kernel exists, and every verdict is recomputable
from them by an outside reader. The oracle therefore provides a standard and clear
check on the speed, correctness and precision of a megakernel.

\input{sec/fig-cells}

\subsection{The end-to-end generation framework}
\label{sec:design:loop}

The ladder says what to build and the oracle what a verdict means; the loop
turns the two into a search. Every round ends on the GPU: a fresh agent edits the
kernel, the kernel is built and measured on an H100, and the gate returns
latency, utilization, profiler counters and error. Each edit follows from a
measurement rather than an estimate, and a cell takes about 48 rounds.

To ensure that successive search rounds are guided by measured evidence rather
than unverified claims, the framework begins each round with a fresh headless
coding agent~\cite{claudecode2025}, as shown in the lower band of
Figure~\ref{fig:arch}. The agent receives the milestone matrix, the current
milestone, the last gate summary and the campaign's rules, and nothing else; the
agent edits the kernel, runs the gate, keeps or reverts the change, appends one
ledger entry and commits. Because a fresh context discards whatever the previous
agent learned but did not record, each entry must carry the hypothesis, the
change, the result and the decision, and only measured values cross a round
boundary.

A gate reads the kernel's behaviour, not the way the round produced that
behaviour, so a few defect classes remain visible only in the diff
(\S\ref{sec:obs:oracle}). After a passing development round, a review round
therefore gives a fresh agent the diff and the run identifiers to audit for
those classes: edits outside the writable surface, a metric obtained by
special-casing a gate's input, and an optimization the preprocessor constants
show was never compiled. A milestone advances only on a gate pass together with
a review pass.

Although passing the gate and the subsequent review establishes
milestone completion, evaluating end-to-end performance requires deployment
within a serving engine to account for its per-token overhead. As an SGLang model
class, the megakernel implements \code{forward(EXTEND)} as a loop of
single-token steps and \code{forward(DECODE)} as the batched step. Both paths
run through the same scheduler, sampler, memory manager and detokenizer on the
same card, so the engine's per-token cost falls inside the measurement
(\S\ref{sec:eval:insertion}).

%% file: sec/fig-arch.tex
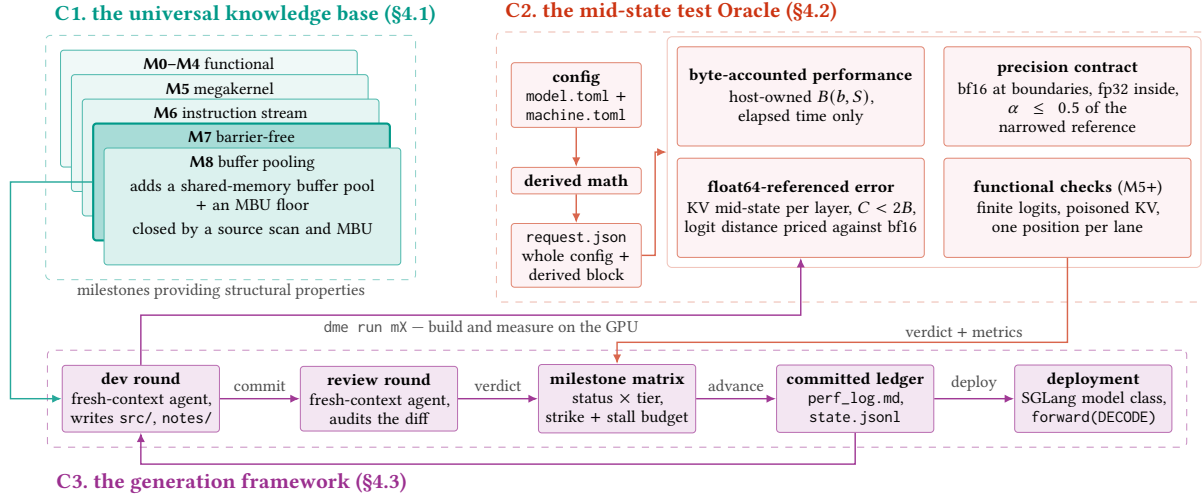
\begin{figure*}[t]
\centering
\small
\scalebox{0.90}{%
\begin{tikzpicture}[
  box/.style={draw=cGrey!80, rounded corners=1.4pt, align=center, inner sep=3pt, font=\small},
  cbox/.style={box, fill=cOurs2!12, draw=cOurs!60},
  mcard/.style={box, draw=cHazy!70, fill=cHazy!8, text width=4.15cm,
                minimum height=1.72cm, font=\scriptsize},
  ocard/.style={cbox, text width=3.42cm, minimum height=1.46cm, font=\scriptsize},
  abox/.style={box, fill=cMpk!12, draw=cMpk!70, text width=2.10cm, font=\scriptsize},
  slbl/.style={font=\scriptsize, text=cGrey},
  ttl/.style={font=\small\bfseries},
  ar/.style={-{Latex[length=4pt,width=3.5pt]}, draw=cOurs!75, line width=0.6pt},
  arg/.style={-{Latex[length=4pt,width=3.5pt]}, draw=cHazy!85, line width=0.6pt},
  aro/.style={-{Latex[length=4pt,width=3.5pt]}, draw=cMpk!95, line width=0.6pt},
]

\foreach \i/\name/\tint in {0/c04/6, 1/c5/8, 2/c6/10}{
  \node[mcard, fill=cHazy!\tint] (\name) at ({2.30+0.16*\i},{0.72-0.36*\i}) {};
}
\node[mcard, draw=cHazy!100, line width=0.9pt, fill=cHazy!34] (c7) at (2.78,-0.36) {};
\node[mcard, fill=cHazy!14] (c8) at (2.94,-0.72) {};
\foreach \name/\head in {c04/{\textbf{\rung{M0}--\rung{M4}} functional},
                         c5/{\textbf{\rung{M5}} megakernel},
                         c6/{\textbf{\rung{M6}} instruction stream},
                         c7/{\textbf{\rung{M7}} barrier-free},
                         c8/{\textbf{\rung{M8}} buffer pooling}}{
  \node[font=\scriptsize, anchor=north, inner sep=0pt] at ($(\name.north)+(0,-0.11)$) {\head};
}
\node[font=\scriptsize, anchor=north, align=center, inner sep=0pt,
      text width=4.15cm] at ($(c8.north)+(0,-0.46)$)
  {adds a shared-memory buffer pool\\ $+$ an MBU floor\\[2pt]
   closed by a source scan and MBU};
\node[slbl, anchor=north west] at (0.25,-1.74)
  {milestones providing structural properties};

\node[cbox, text width=1.70cm, font=\scriptsize] (cfg)   at (7.70, 0.92)
  {\textbf{config}\\ \texttt{\scriptsize{}model.toml} $+$\\ \texttt{\scriptsize{}machine.toml}};
\node[cbox, text width=1.70cm, font=\scriptsize] (mrg)   at (7.70,-0.34)
  {\textbf{derived math}};
\node[cbox, text width=1.70cm, font=\scriptsize] (req)   at (7.70,-1.44)
  {\texttt{\scriptsize{}request.json}\\ whole config $+$\\ derived block};
\draw[ar] (cfg.south) -- (mrg.north);
\draw[ar] (mrg.south) -- (req.north);

\node[ocard] (obytes) at (11.00, 0.90) {\textbf{byte-accounted performance}\\[1pt]
  host-owned $B(b,S)$, elapsed time only};
\node[ocard] (oprec)  at (14.92, 0.90) {\textbf{precision contract}\\[1pt]
  bf16 at boundaries, fp32 inside,\\ $\alpha\le0.5$ of the narrowed reference};
\node[ocard] (oerr)   at (11.00,-0.74) {\textbf{float64-referenced error}\\[1pt]
  KV mid-state per layer, $C<2B$,\\ logit distance priced against bf16};
\node[ocard] (ofunc)  at (14.92,-0.74) {\textbf{functional checks} (\rung{M5}$+$)\\[1pt]
  finite logits, poisoned KV,\\ one position per lane};
\node[draw=cOurs!30, rounded corners=2pt, fit=(obytes)(oerr)(oprec)(ofunc),
      inner sep=4pt] (checks) {};
\draw[ar] (req.east) -- (8.86,-1.44) -- (8.86, 0.08) -- (checks.west |- 0,0.08);

\node[abox] (dev) at (1.30,-3.55)
  {\textbf{dev round}\\ fresh-context agent,\\ writes \texttt{\scriptsize{}src/}, \texttt{\scriptsize{}notes/}};
\node[abox] (rev) at (4.80,-3.55)
  {\textbf{review round}\\ fresh-context agent,\\ audits the diff};
\node[abox] (mat) at (8.30,-3.55)
  {\textbf{milestone matrix}\\ status $\times$ tier,\\ strike $+$ stall budget};
\node[abox] (led) at (11.80,-3.55)
  {\textbf{committed ledger}\\ \texttt{\scriptsize{}perf\_log.md},\\ \texttt{\scriptsize{}state.jsonl}};
\node[abox] (dep) at (15.30,-3.55)
  {\textbf{deployment}\\ SGLang model class,\\ \texttt{\scriptsize{}forward(DECODE)}};

\draw[aro] (dev) -- node[slbl,above]{commit} (rev);
\draw[aro] (rev) -- node[slbl,above]{verdict} (mat);
\draw[aro] (mat) -- node[slbl,above]{advance} (led);
\draw[aro] (led) -- node[slbl,above]{deploy} (dep);
\draw[aro] (led.south) -- ++(0,-0.46) -| (dev.south);

\draw[aro] (dev.north) -- (1.30,-2.32) -- (11.00,-2.32) -- (oerr.south);
\node[slbl, below, inner sep=2pt] at (6.30,-2.32)
  {\texttt{\scriptsize{}dme run mX} --- build and measure on the GPU};
\draw[ar] (ofunc.south) -- (14.92,-2.72) -- (8.30,-2.72) -- (mat.north);
\node[slbl, above, inner sep=2pt] at (13.40,-2.72) {verdict $+$ metrics};

\draw[arg] (c7.west) -- ++(-1.20,0) |- (dev.west);

\begin{scope}[on background layer]
  \node[draw=cHazy!45, dashed, rounded corners=2pt, fit=(c04)(c8), inner sep=6pt] (f2) {};
  \node[draw=cOurs!35, dashed, rounded corners=2pt,
        fit=(cfg)(req)(obytes)(oprec)(ofunc), inner sep=6pt] (f1) {};
  \node[draw=cMpk!55, dashed, rounded corners=2pt, fit=(dev)(dep), inner sep=6pt] (f3) {};
\end{scope}
\node[ttl, anchor=south west, text=cHazy] at ($(f2.north west)+(0.03,0.03)$)
  {C1.\ the universal knowledge base (\S\ref{sec:design:ladder})};
\node[ttl, anchor=south west, text=cOurs] at ($(f1.north west)+(0.03,0.03)$)
  {C2.\ the mid-state test Oracle (\S\ref{sec:design:oracle})};
\node[ttl, anchor=north west, text=cMpk] at ($(f3.south west)+(0.03,-0.26)$)
  {C3.\ the generation framework (\S\ref{sec:design:loop})};

\end{tikzpicture}}
\caption{\sys{} overview. C1 states the structure a kernel must reach, one card
per milestone. C2 derives every quantity a verdict rests on. C3 runs one
fresh-context round per edit and deploys the kernel.}
\label{fig:arch}
\end{figure*}

%% file: sec/tab-ladder.tex
\begin{table}[t]
\centering
\footnotesize
\setlength{\tabcolsep}{4pt}
\caption{The ten milestones. Each milestone adds one property to the megakernel
that passes the previous milestone.}
\label{tab:ladder}
\begin{tabular}{@{}ll>{\raggedright\arraybackslash}p{0.30\columnwidth}l@{}}
\toprule
& \textbf{name} & \textbf{what it adds} & \textbf{checked by} \\
\midrule
\rung{M0}--\rung{M4} & functional & toolchain, operators, a layer, the stack,
decode & float64 golden \\
\rung{M5} & megakernel & one persistent launch, no library calls & launch
count \\
\rung{M6} & instruction stream & a per-SM instruction stream $+$ MBU floor &
schedule audit \\
\rung{M7} & barrier-free & counter dependencies $+$ MBU floor & scan, timeline
\\
\rung{M8} & buffer pooling & a shared-memory buffer pool $+$ MBU floor & source scan \\
\rung{M9} & stability & a 1000-step run (optional) & greedy drift \\
\bottomrule
\end{tabular}
\end{table}

%% file: sec/fig-gate.tex
\begin{algorithm}[t]
\caption{The oracle gate design.}
\label{alg:gate}
\newcommand{\gc}[1]{\normalfont\itshape\textcolor{cGrey}{#1}\ttfamily}
\newcommand{\gi}{\hspace*{1.15em}}
{\ttfamily\footnotesize\raggedright
\textbf{gate(m, kernel):}\\
\gi d = derive(model\_cfg, machine\_cfg)\\
\gi\gi \gc{\# B(b,S), tolerances, D, floors}\\[2pt]
\gi \gc{\# 1. numerics, on the kernel this milestone built}\\
\gi \textbf{if} m == M1: \gc{\# per op, on the oracle's taps}\\
\gi\gi alpha(kernel.op, A, A') $\le$ 0.5 \gc{\# width, \eqcite{eq:alpha}}\\
\gi\gi \gc{\# A == A': abstain, never pass}\\
\gi \textbf{if} m $\ge$ M5:\\
\gi\gi finite(logits) \gc{\# no NaN or infinity}\\
\gi\gi logits == logits(poison(kv)) \gc{\# OOB KV read}\\
\gi\gi decode(lane, pos[lane]) \gc{\# no lane-0 broadcast}\\
\gi\gi C(kernel, torch) < 2$\cdot$B(torch, fp64) \gc{\# \eqcite{eq:cb}}\\
\gi\gi err $\le$ 5$\cdot$d.D, err $\le$ err(sgl), \gc{\# the three bars}\\
\gi\gi\gi\gi\gi\gi\gi err\_max $\le$ 2$\cdot$err\_max(sgl)\\
\gi\gi \gc{\# no fp64 family: abstain, never pass}\\[2pt]
\gi \gc{\# 2. structure, by source scan and schedule audit}\\
\gi \textbf{if} m == M5: launches/step $\le$ 1, no libcalls\\
\gi \textbf{if} m == M6: types $\ge$ 5, instr./layer $\ge$ 6,\\
\gi\gi\gi\gi\gi\gi\gi imbalance $\le$ 1.35\\
\gi \textbf{if} m == M7: barriers == 0, counters $\ge$ 4L,\\
\gi\gi\gi\gi\gi\gi\gi busy\_frac $\ge$ 0.85, tail\_spread $\le$ 0.05\\
\gi \textbf{if} m == M8: shared-memory buffer pool present\\
\gi \textbf{if} m == M9: drift(1000 steps) $\le$ 5\\[2pt]
\gi \gc{\# 3. the number; the kernel returns only us}\\
\gi us = (kernel.decode(80) $-$ kernel.decode(16))/64\\
\gi \gc{\# kernel.scope declared, never mixed}\\
\gi mbu = d.B / (us $\cdot$ d.beta\_peak) \gc{\# from the config}\\
\gi mbu $\le$ 100, mbu $\le$ 1/rho(ncu) \gc{\# bytes actually read}\\
\gi \textbf{if} m $\in$ \{M6, M7, M8\}: mbu $\ge$ d.floor[m]\\[2pt]
\gi \textbf{return} \{status, suite, summary, metrics, details\}\par}
\end{algorithm}

%% file: sec/fig-cells.tex
\begin{figure*}[t]
\centering
\includegraphics[width=\textwidth]{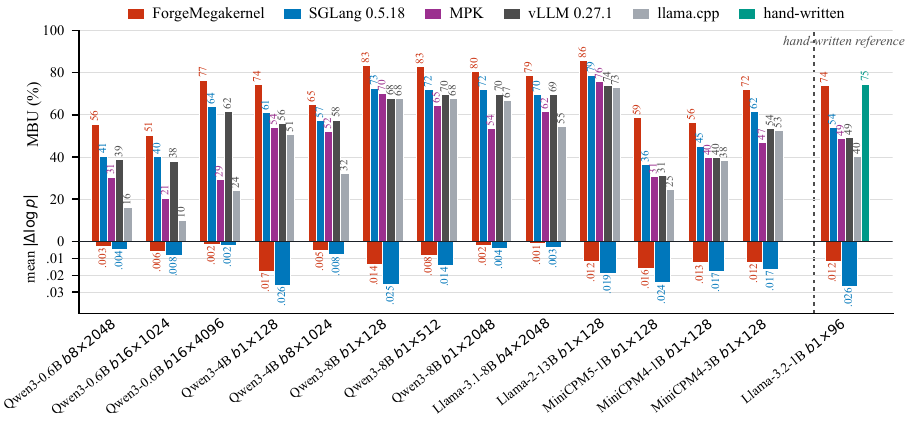}
\caption{The 14 forged megakernels against four baselines: isolated-step MBU
(upward) and log-probability error against a float64 forward pass (downward).}
\label{fig:cells}
\end{figure*}

%% file: sec/05-evaluation.tex
\section{Evaluation}
\label{sec:eval}

\input{sec/fig-step}

\subsection{Setup}
\label{sec:eval:setup}

\begin{sloppypar}
Measurements use one NVIDIA H100 80GB HBM3 at 1980\,MHz with
$\beta_{\text{peak}} = 3350$\,GB/s. Each run uses one GPU, bf16 weights and KV
cache. Runs on clock-locked cards are excluded rather
than rescaled.
We compute every latency as the two-point slope defined in
\S\ref{sec:design:oracle}. All reported measurements use the end-to-end clock,
which is also the clock used by SGLang and MPK.
\end{sloppypar}

\begin{sloppypar}
\textbf{Baselines.} We compare against SGLang~0.5.18~\cite{zheng2024sglang}
with CUDA graphs and \code{torch.compile} enabled, vLLM~0.27.1
~\cite{kwon2023pagedattention} with CUDA graphs enabled and
\code{detokenize=False}, llama.cpp~\cite{llamacpp} with bf16 weights in its
GGUF format, all layers resident, and flash attention, and the Mirage Persistent
Kernel (MPK)~\cite{mpk2025}, executed through its demo path. We measure each
engine at its newest release, tuned to the best settings we could find.
\end{sloppypar}

\textbf{Models and configurations.} We run 14 representative cells drawn from
eight model families:
Llama-3.2-1B~\cite{grattafiori2024llama3}, Llama-3.1-8B,
Llama-2-13B-chat~\cite{touvron2023llama2}, Qwen3-0.6B/1.7B/4B/8B
~\cite{yang2025qwen3}, Qwen2.5-3B, MiniCPM4-0.5B/1B/3B, MiniCPM5-1B
~\cite{hu2024minicpm}, and NanBeige4-3B. A configuration denotes one $(b,S)$ pair,
with batch sizes from 1 to 16 and contexts from 96 to 4096 tokens.

\textbf{Agent.} Each round uses the same headless coding agent, Claude Code
~\cite{claudecode2025} running Claude Opus 5 with a 1M-token context window,
with a fresh context and access only to a shell, the repository, and the
\sys{} command.

\subsection{Kernel experiment}
\label{sec:eval:kernel}
\label{sec:eval:forge}


\subsubsection{Main results}
\label{sec:eval:main}

Figure~\ref{fig:cells} compares the 14 forged megakernels with four baselines
at the same configurations. The Llama-3.2-1B $b1{\times}96$ kernel is forged
against the hand-written reference and is excluded from the aggregates below.
The forged megakernels reach 50.5--85.9\% MBU, whereas SGLang reaches
36.5--78.9\% and MPK 20.6--75.9\%; smaller models have less matrix work to keep
each SM busy, so their MBU is lower. Our method improves on SGLang by
$1.10$--$1.61\times$ (geometric mean $1.21\times$) and on MPK by
$1.14$--$2.51\times$ (geometric mean $1.54\times$).

On Llama-3.2-1B, a hand-written megakernel~\cite{spector2025nobubbles} reaches
74.6\% MBU (our own measurement of the reproduced kernel on our card; its
authors report 78\% on theirs), and the forged
kernel reaches a comparable 74.0\% on the same configuration---a gain of $1.37\times$,
$1.50\times$, $1.52\times$, and $1.84\times$ over SGLang, vLLM, MPK, and
llama.cpp, respectively.

\subsubsection{\draftname{Correctness}}
\label{sec:eval:acc}

We check each kernel against a float64 forward pass of the same checkpoint.
Both forward passes are teacher-forced on the same prompts.

\textbf{Log-probability error.} We take the value a decoder assigns to the token
that actually follows, in log space, and compare it with the float64 value at
the same position. The downward bars of Figure~\ref{fig:cells} report the mean
absolute difference, in nats, over all positions of a configuration.

The forged megakernels are closer to exact arithmetic than SGLang on every
configuration. The ratio of their mean error to SGLang's ranges from 0.38
to 0.72, with a median of 0.66.

\subsubsection{Decomposing a decode step}
\label{sec:eval:profile}

\input{sec/prof-body}

\input{sec/tab-gsm8k}
\input{sec/seq-figure}
\input{sec/ex-figure}

\subsection{Serving experiment}
\label{sec:eval:serving}
\label{sec:eval:insertion}
\label{sec:eval:gsm8k}
\label{sec:eval:example}
\label{sec:eval:seqlen}

We evaluate the generated megakernels inside a production engine on the GSM8K
workload. Each megakernel is deployed as an SGLang model class and executed through
\code{Engine.generate} (\S\ref{sec:design:loop}), so the generated megakernel and the
unmodified baseline use the same serving stack: scheduler, sampler, memory
manager, and detokenizer. CUDA graphs are enabled for the baseline and disabled
for the generated megakernel. The workload
is GSM8K~\cite{cobbe2021gsm8k}: 200 test questions with ragged prompt lengths,
identical for both. We extract the predicted number from each answer and
compare it numerically with the gold answer.

Table~\ref{tab:gsm8k} reports the results for the 13 generated megakernels.
The speedups range from
$1.03\times$ to $1.28\times$ with a geometric mean of $1.11\times$. On
accuracy, the forged megakernels stay within 1.5 percentage points of the baseline
and match or exceed it on seven configurations. The difference falls within
run-to-run noise: the baseline itself varies by the same margin across repeated
runs, because ragged batches are scheduled differently.

Figure~\ref{fig:example} shows one GSM8K question with the full generated
answer from four engines on the same prompt and greedy decoding: the SGLang engine,
vLLM, llama.cpp, and the generated megakernel. All four produce the correct answer,
and the generated megakernel has the lowest measured per-token latency on this
question: $1.04\times$ SGLang,
$1.16\times$ vLLM and $1.26\times$ llama.cpp. The generated texts differ
token for token, yet every answer is correct; the same holds across
the six questions we generated this way.

Each megakernel fixes its batch size and context bound at compile time, so the
deployment retains several kernels and routes requests by context length.
Figure~\ref{fig:seq} sweeps four Qwen3-8B $b{=}1$ kernels, generated from one seed
at four context bins, across $S \in \{64, \ldots, 4096\}$. The upper envelope
of the four kernels determines the best measured MBU at each length, and the
shading indicates the selected kernel. The kernel generated at a larger bin is
generally the better choice, and only the
$S_f{=}4096$ kernel can serve $S{=}4096$. The selected kernel set reaches
79.6--83.7\% MBU
over $S=64$--4096, against SGLang's 72.8--74.1\%.

%% file: sec/fig-step.tex
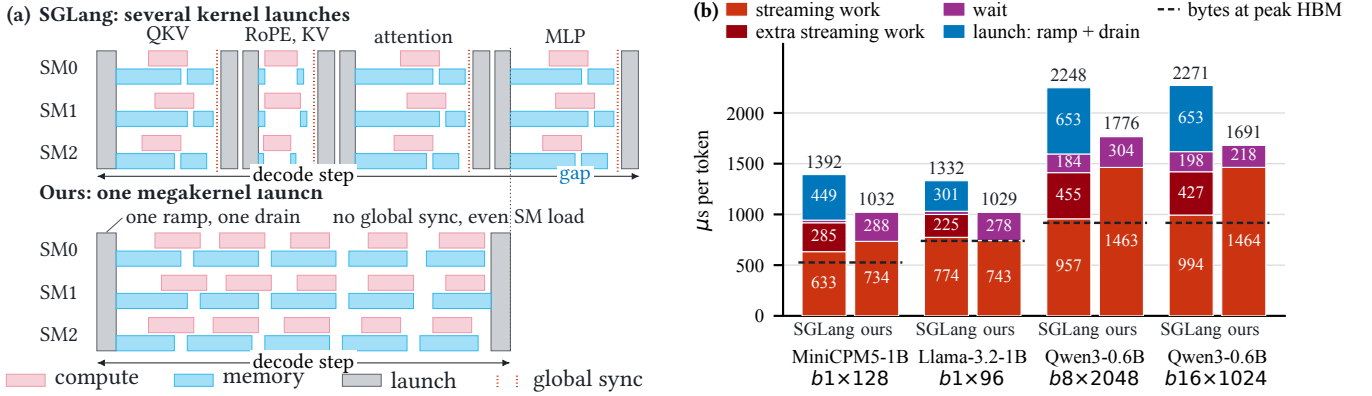
\begin{figure*}[t]
\centering
\begin{minipage}[b]{0.485\textwidth}
\centering
\begin{tikzpicture}[
  x=0.0855cm, y=1cm,
  cbox/.style={draw=cOurs2, fill=cOurs2!45, line width=0.25pt},
  mbox/.style={draw=cAccent, fill=cAccent!45, line width=0.25pt},
  lbox/.style={draw=cInk!70, fill=cMute!55,  line width=0.25pt},
  synl/.style={draw=cOurs, line width=0.75pt, dash pattern=on 0.4pt off 0.9pt},
  cut/.style={draw=cGrey, line width=0.4pt, dash pattern=on 0.5pt off 0.9pt},
  lead/.style={draw=cGrey, line width=0.3pt},
  ann/.style={font=\footnotesize, text=cInk, inner sep=1pt},
  knm/.style={font=\footnotesize, text=cInk, inner sep=1pt},
  ttl/.style={font=\footnotesize\bfseries, anchor=west, inner sep=0pt},
  sml/.style={font=\footnotesize, text=cInk, anchor=east, inner sep=2pt},
  meas/.style={{Latex[length=3pt,width=2.6pt]}-{Latex[length=3pt,width=2.6pt]},
               draw=cInk, line width=0.35pt},
  tag/.style={font=\footnotesize, inner sep=1pt, fill=white},
  lgd/.style={font=\small, text=cInk, inner sep=1pt},
]
\def\Cbx#1#2#3{\draw[cbox] (#1,#3-0.1) rectangle (#2,#3+0.1);}
\def\Mbx#1#2#3{\draw[mbox] (#1,#3-0.34) rectangle (#2,#3-0.14);}

\node[ttl, text=cInk] at (-14,0.56) {\textbf{(a)}\, SGLang: several kernel launches};

\foreach \a/\b in {0/3, 22.6/25, 37.6/40, 61.6/64}       
  {\draw[lbox] (\a,-1.46) rectangle (\b,0.10);}
\foreach \a/\b in {19.2/21.8, 34.2/36.8, 58.2/60.8, 81.2/83.8} 
  {\draw[lbox] (\a,-1.46) rectangle (\b,0.10);}
\foreach \s in {18.6, 33.6, 57.6, 80.6} {\draw[synl] (\s,-1.50) -- (\s,0.14);}

\node[knm] at (11,0.30)   {QKV};
\node[knm] at (29.5,0.30) {RoPE, KV};
\node[knm] at (49,0.30)   {attention};
\node[knm] at (72.5,0.30) {MLP};

\Mbx{3}{13}{0}      \Mbx{14}{18}{0}      \Cbx{8}{14}{0}
\Mbx{3}{14}{-0.56}  \Mbx{15}{18}{-0.56}  \Cbx{9}{15}{-0.56}
\Mbx{3}{12}{-1.12}  \Mbx{13}{17}{-1.12}  \Cbx{7}{13}{-1.12}
\Mbx{25}{26}{0}       \Mbx{31}{32}{0}       \Cbx{26}{31}{0}
\Mbx{25}{26}{-0.56}   \Mbx{31.5}{32.5}{-0.56} \Cbx{26}{31.5}{-0.56}
\Mbx{25}{25.8}{-1.12} \Mbx{30}{30.8}{-1.12} \Cbx{25.8}{30}{-1.12}
\Mbx{40}{52}{0}     \Mbx{53}{57}{0}     \Cbx{47}{53}{0}
\Mbx{40}{53}{-0.56} \Mbx{54}{57}{-0.56} \Cbx{48}{54}{-0.56}
\Mbx{40}{51}{-1.12} \Mbx{52}{56}{-1.12} \Cbx{46}{52}{-1.12}
\Mbx{64}{75}{0}     \Mbx{76}{80}{0}     \Cbx{70}{76}{0}
\Mbx{64}{76}{-0.56} \Mbx{77}{80}{-0.56} \Cbx{71}{77}{-0.56}
\Mbx{64}{74}{-1.12} \Mbx{75}{79}{-1.12} \Cbx{69}{75}{-1.12}

\foreach \y/\n in {-0.12/SM0, -0.68/SM1, -1.24/SM2} {\node[sml] at (-2,\y) {\n};}

\node[ttl, text=cInk] at (-14,-1.80) {\textbf{\phantom{(a)}}\, Ours: one megakernel launch};

\draw[lbox] (0,-3.87)  rectangle (3,-2.31);
\draw[lbox] (61,-3.87) rectangle (64,-2.31);

\foreach \a/\b in {3/13, 15/25, 27/37, 39/48, 51/60} {\Mbx{\a}{\b}{-2.41}}
\foreach \a/\b in {9/16, 19/26, 30/36, 42/48, 53/60} {\Cbx{\a}{\b}{-2.41}}
\foreach \a/\b in {3/14, 16/25, 27/37, 39/49, 52/61} {\Mbx{\a}{\b}{-2.97}}
\foreach \a/\b in {10/17, 20/27, 31/37, 42/49, 54/60} {\Cbx{\a}{\b}{-2.97}}
\foreach \a/\b in {3/12, 14/24, 26/36, 38/48, 50/59} {\Mbx{\a}{\b}{-3.53}}
\foreach \a/\b in {8/15, 18/25, 29/36, 41/47, 52/58} {\Cbx{\a}{\b}{-3.53}}

\foreach \y/\n in {-2.53/SM0, -3.09/SM1, -3.65/SM2} {\node[sml] at (-2,\y) {\n};}

\node[ann] (lchB) at (18,-2.11) {one ramp, one drain};
\draw[lead] (lchB.west) -- (1.5,-2.30);
\node[ann] at (56,-2.11) {no global sync, even SM load};

\draw[cut] (64,0.14) -- (64,-3.89);
\draw[meas] (0,-1.56) -- (84,-1.56);
\node[tag] at (32,-1.56) {decode step};
\node[tag, text=cSglang] at (74,-1.56) {gap};
\draw[meas] (0,-4.03) -- (64,-4.03);
\node[tag] at (32,-4.03) {decode step};

\def\lgy{-4.26}
\draw[cbox] (-14,\lgy-0.08) rectangle (-8,\lgy+0.08);
\node[lgd, anchor=west] at (-7,\lgy) {compute};
\draw[mbox] (12,\lgy-0.08) rectangle (18,\lgy+0.08);
\node[lgd, anchor=west] at (19,\lgy) {memory};
\draw[lbox] (38,\lgy-0.08) rectangle (44,\lgy+0.08);
\node[lgd, anchor=west] at (45,\lgy) {launch};
\draw[synl] (62,\lgy-0.09) -- (62,\lgy+0.09);
\draw[synl] (65,\lgy-0.09) -- (65,\lgy+0.09);
\node[lgd, anchor=west] at (67,\lgy) {global sync};
\end{tikzpicture}
\end{minipage}\hfill
\begin{minipage}[b]{0.485\textwidth}
\centering
\begin{tikzpicture}[inner sep=0pt, outer sep=0pt]
  \node[anchor=south west] (bimg)
    {\includegraphics[width=\linewidth]{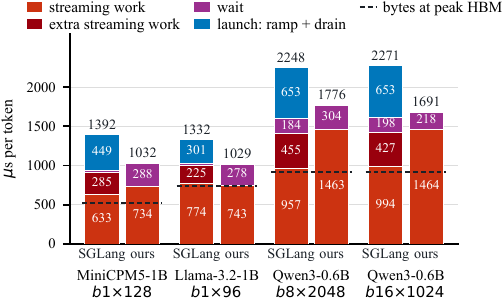}};
  \node[font=\footnotesize\bfseries, anchor=north west, overlay]
    at ($(bimg.north west)+(0.02,-0.02)$) {(b)};
\end{tikzpicture}
\end{minipage}
\caption{Decoding latency per SM, the terms of \eqcite{eq:step}.
\textbf{(a)} the two engines drawn schematically: grey is the launch cost
$\ell$, which SGLang repeats with a grid-wide sync per kernel and ours incurs
once. \textbf{(b)} a decode step's measured microseconds on four cells, split
into the same terms; the dashed line is what \eqcite{eq:bytes} alone would take
at $\beta_{\text{peak}}$.}
\label{fig:gantt}
\label{fig:why}
\end{figure*}

%% file: sec/prof-body.tex
We profile four configurations to locate the forged megakernels' advantage. A
decode step is a sequence of grid launches, where a launch is the device-side
start and retirement of a grid rather than a host call: graph replay removes the
host cost and leaves this one. For the launches $\mathcal{K}$ of one step and one
SM $i$,
\begin{equation}
\label{eq:step}
t \;=\; \sum_{k \in \mathcal{K}} \Big( \ell_k \;+\; \max_{i}\,\big[\, s_{i,k} + w_{i,k} \,\big] \Big),
\end{equation}
where $\ell$ is the grid's own ramp and drain, $s$ is the time an SM spends
streaming bytes and computing on them, and $w$ is the time it spends waiting
for another SM. Figure~\ref{fig:gantt}(a) draws these terms for both engines. The
SGLang backend runs one grid per operator, so it repeats $\ell$ at every launch
and $w$ at every launch boundary. Ours runs one grid per step, so it incurs
$\ell$ once and waits only where its own counters require.
Figure~\ref{fig:why}(b) splits each engine's measured step into the same slots, so
the difference between the two bars is the gap.

\textbf{Launch time.} The largest term is $\ell$, the grey bands in the
SGLang panel of Figure~\ref{fig:gantt}(a). The profiled SGLang runs replay a
captured CUDA graph, which removes the host cost (the CPU work of issuing each
kernel), but still starts and retires a grid per kernel on the GPU, and the ramp
and drain of each grid cause the launch latency $\ell$. We model each SGLang
matvec grid latency as $t_k = c + B_k/\beta_{\text{peak}}$ for its bytes $B_k$, regressing the measured durations and giving $c = 4.63$\us{} per
launch ($R^2 = 0.9993$), or 301--653\us{} over the 65--141
GEMV-family launches per
token, which covers 99--138\% of every gap in Figure~\ref{fig:why}(b). The megakernel launches once per step and
costs only 9.3\us{}.

\textbf{Kernel sync.} A launch boundary is also a grid-wide sync: no SM starts
the next kernel until every SM has arrived, so the slowest sets the boundary
and the rest wait. Counted this way, with a ramp and a drain recorded as SMs
waiting, SGLang waits 333--851\us{} per token and the megakernel
227--313\us{}, of which 218--304\us{} is spent at dependency counters rather
than at launch boundaries.

\textbf{Fusion.} Splitting the step with multiple launches also forces 111--239
launches per token that carry almost no model bytes, including standalone
RMSNorm, RoPE, the KV store and the attention-combine. They cost
225--455\us{}, 17--21\% of the step, while the megakernel performs the same arithmetic in the next
instruction of the same kernel, on operands that are still in registers.
\eqcite{eq:bytes} counts reads, and both engines read the same bytes; they
differ in their writes. A split step passes each intermediate to the next kernel
through DRAM, and \code{ncu} measures SGLang writing 7.7\,MB per token on
MiniCPM5-1B and 67--168\,MB on Qwen3-0.6B at $b8{\times}2048$, against 1.4 and
2.3\,MB for the megakernel. At $\beta_{\text{peak}}$ those differences cost 0.8
and 43\us{}, far less than the launch and sync time the same kernels incur.

%% file: sec/tab-gsm8k.tex
\begin{table}[t]
\centering
\footnotesize
\setlength{\tabcolsep}{3.2pt}
\caption{GSM8K inside SGLang: answer accuracy and latency per decoded token.}
\label{tab:gsm8k}
\begin{tabular}{@{}lrrrrrr@{}}
\toprule
& & & \multicolumn{2}{c}{latency (\us{}/tok)} & \multicolumn{2}{c}{accuracy} \\
\cmidrule(lr){4-5}\cmidrule(lr){6-7}
Model & $b$ & $S$ & ours & baseline & ours & baseline \\
\midrule
Qwen3-0.6B & 8 & 2048 & \textbf{1406} & 1576 & 0.560 & 0.555 \\
Qwen3-0.6B & 16 & 1024 & \textbf{1457} & 1709 & 0.555 & 0.555 \\
Qwen3-0.6B & 16 & 4096 & \textbf{1322} & 1688 & 0.545 & 0.550 \\
Qwen3-4B & 1 & 128 & \textbf{3387} & 3804 & 0.935 & 0.930 \\
Qwen3-4B & 8 & 1024 & \textbf{4217} & 4459 & 0.940 & 0.930 \\
Qwen3-8B & 1 & 128 & \textbf{5601} & 6009 & 0.965 & 0.975 \\
Qwen3-8B & 1 & 512 & \textbf{5757} & 6116 & 0.965 & 0.970 \\
Qwen3-8B & 1 & 2048 & \textbf{5847} & 6219 & 0.970 & 0.970 \\
\addlinespace
Llama-3.1-8B & 4 & 2048 & \textbf{6130} & 6378 & 0.880 & 0.865 \\
Llama-2-13B-chat & 1 & 128 & \textbf{9498} & 9761 & 0.315 & 0.315 \\
\addlinespace
MiniCPM5-1B & 1 & 128 & \textbf{1181} & 1430 & 0.550 & 0.555 \\
MiniCPM4-1B & 1 & 128 & \textbf{1718} & 1970 & 0.640 & 0.650 \\
MiniCPM4-3B & 1 & 128 & \textbf{3145} & 3394 & 0.620 & 0.635 \\
\bottomrule
\end{tabular}
\end{table}

%% file: sec/seq-figure.tex
\begin{figure}[t]
\centering
\includegraphics[width=\columnwidth]{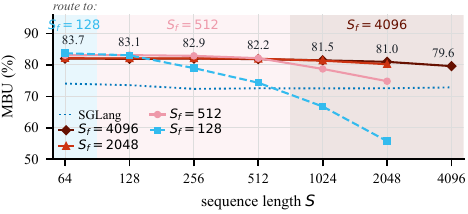}
\caption{Sequence-length sweep of four Qwen3-8B $b{=}1$ forged megakernels.}
\label{fig:seq}
\end{figure}

%% file: sec/ex-figure.tex
\begin{figure}[t]
\footnotesize
\setlength{\fboxsep}{2.5pt}
\noindent\fbox{\parbox{0.955\columnwidth}{%
\textbf{Prompt} (Qwen3-4B, greedy, after a fixed \emph{solve step by
step, end with} \code{Answer: \textbackslash{}boxed\{\$Answer\}}
preamble)\textbf{.} Ryan plants 2 flowers a day in his garden. After 15 days, how many flowers
does he have if 5 did not grow? \hfill\emph{gold:} 25
}}

\vspace{1.5pt}
\noindent\fbox{\parbox{0.955\columnwidth}{%
\textbf{llama.cpp} \hfill 4768\us/tok \quad 50.5\% MBU \quad 87 tok \quad \textbf{25} \checkmark

\vspace{1pt}
{\ttfamily\scriptsize Ryan plants 2 flowers each day for 15 days. So, the total number of
flowers planted is: \$\$ 2 \textbackslash{}times 15 = 30 \$\$ However, 5
of the flowers did not grow. So, the number of flowers that actually grew
is: \$\$ 30 - 5 = 25 \$\$ Answer: \textbackslash{}boxed\{25\}}
}}

\vspace{1.5pt}
\noindent\fbox{\parbox{0.955\columnwidth}{%
\textbf{vLLM 0.27.1} \hfill 4382\us/tok \quad 54.9\% MBU \quad 102 tok \quad \textbf{25} \checkmark

\vspace{1pt}
{\ttfamily\scriptsize Ryan plants 2 flowers each day for 15 days. So, the total number of
flowers planted is: \$\$ 2 \textbackslash{}text\{ flowers/day\}
\textbackslash{}times 15 \textbackslash{}text\{ days\} = 30
\textbackslash{}text\{ flowers\} \$\$ However, 5 of the flowers did not
grow. So, the number of flowers that actually grew is: \$\$ 30 - 5 = 25
\$\$ Answer: \textbackslash{}boxed\{25\}}
}}

\vspace{1.5pt}
\noindent\fbox{\parbox{0.955\columnwidth}{%
\textbf{SGLang 0.5.18} \hfill 3932\us/tok \quad 61.2\% MBU \quad 87 tok \quad \textbf{25} \checkmark

\vspace{1pt}
{\ttfamily\scriptsize Ryan plants 2 flowers each day for 15 days. So, the total number of
flowers planted is: \$\$ 2 \textbackslash{}times 15 = 30 \$\$ However, 5
of the flowers did not grow. So, the number of flowers that actually grew
is: \$\$ 30 - 5 = 25 \$\$ Answer: \textbackslash{}boxed\{25\}}
}}

\vspace{1.5pt}
\noindent\fbox{\parbox{0.955\columnwidth}{%
\textbf{\sys{}} \hfill \textbf{3781}\us/tok \quad \textbf{63.7\%} MBU \quad 102 tok \quad \textbf{25} \checkmark

\vspace{1pt}
{\ttfamily\scriptsize Ryan plants 2 flowers each day for 15 days. So, the total number of
flowers planted is: \$\$ 2 \textbackslash{}text\{ flowers/day\}
\textbackslash{}times 15 \textbackslash{}text\{ days\} = 30
\textbackslash{}text\{ flowers\} \$\$ However, 5 of the flowers did not
grow. So, the number of flowers that actually grew is: \$\$ 30 - 5 = 25
\$\$ Answer: \textbackslash{}boxed\{25\}}
}}

\caption{One GSM8K question and the answer from four engines on the same
prompt.}
\label{fig:example}
\end{figure}

%% file: sec/06-discussion.tex
\section{Discussion and Limitations}
\label{sec:disc}

Our evaluation runs on one GPU architecture (an H100 80GB, 132 SMs, HBM3 at
$\beta_{\text{peak}} = 3350$\,GB/s). Whether the same milestones hold on different hardware,
with another shared-memory budget, core count or launch model, is open, and
answering it calls for further experiments across architectures and
configurations.

\input{sec/tab-cost}
Table~\ref{tab:cost} reports the resources consumed while forging each megakernel,
summed from the per-round envelopes that record wall-clock time, token usage, and
billed cost. The median megakernel required 47 rounds, 14.7 agent-hours and
178\,Mtok, or \$177. For large-scale or long-running serving deployments, this
cost is readily justified by the efficiency gains delivered by the megakernel.

%% file: sec/tab-cost.tex
\begin{table}[t]
\centering
\footnotesize
\setlength{\tabcolsep}{1.1pt}
\caption{Cost of forging each megakernel.}
\label{tab:cost}
\begin{tabular}{@{}lrrrrrrrrr@{}}
\toprule
& & & & & & & \multicolumn{2}{c}{tokens (M)} & \\
\cmidrule(lr){8-9}
Model & $b$ & $S$ & rounds & gates & agent~h & span~d & prompt & out & cost \\
\midrule
Qwen3-0.6B & 8 & 2048 & 40 & 62 & 22.7 & 1.0 & 176.0 & 1.91 & 177 \\
Qwen3-0.6B & 16 & 1024 & 76 & 90 & 21.1 & 12.2 & 295.8 & 3.11 & 307 \\
Qwen3-0.6B & 16 & 4096 & 71 & 59 & 17.6 & 9.0 & 304.3 & 2.29 & 275 \\
Qwen3-4B & 1 & 128 & 102 & 77 & 25.1 & 2.6 & 339.6 & 3.70 & 366 \\
Qwen3-4B & 8 & 1024 & 58 & 62 & 40.4 & 11.1 & 330.9 & 3.49 & 335 \\
Qwen3-8B & 1 & 128 & 37 & 32 & 6.1 & 8.4 & 114.5 & 1.02 & 109 \\
Qwen3-8B & 1 & 512 & 29 & 27 & 8.6 & 1.2 & 131.7 & 1.15 & 127 \\
Qwen3-8B & 1 & 2048 & 23 & 16 & 7.1 & 1.2 & 84.0 & 0.57 & 77 \\
Llama-3.1-8B & 4 & 2048 & 69 & 57 & 22.4 & 8.9 & 414.2 & 2.90 & 362 \\
Llama-2-13B-chat & 1 & 128 & 27 & 21 & 5.3 & 8.4 & 95.5 & 0.65 & 83 \\
MiniCPM5-1B & 1 & 128 & 37 & 28 & 5.8 & 0.3 & 97.0 & 1.03 & 104 \\
MiniCPM4-1B & 1 & 128 & 60 & 57 & 14.7 & 13.2 & 195.3 & 2.30 & 214 \\
MiniCPM4-3B & 1 & 128 & 47 & 42 & 8.9 & 14.4 & 136.3 & 1.21 & 132 \\
Llama-3.2-1B & 1 & 96 & 77 & 92 & 27.1 & 2.7 & 401.6 & 4.46 & 414 \\
\midrule
\textit{median} & & & 47 & 57 & 14.7 & 8.4 & 176.0 & 1.91 & 177 \\
\bottomrule
\end{tabular}
\end{table}

%% file: sec/07-related.tex
\section{Related Work}
\label{sec:related}

\paragraph{Hand-written megakernels.} Two recent systems implement the complete
forward pass as a megakernel: a Llama-1B implementation built around an on-GPU
interpreter, reporting 78\% of H100 bandwidth~\cite{spector2025nobubbles} and
reproduced at 74.6\% on our card, and FlashFormer, built around a shared
pipelined buffer~\cite{nrusimha2025flashformer}. Both target one model and one
configuration per kernel, as do extensions to multiple
GPUs~\cite{juravsky2025wholegpu} and other vendors~\cite{kog2026monokernel}. The
ThunderKittens authors subsequently identified the need for an oracle independent
of the generated layer~\cite{retireabstractions}. \rev{Their validation relies on
expert-defined checks per kernel; \sys{} derives the measurement quantities from
each model configuration.}

\paragraph{Compiled megakernels.} MPK lowers tensor programs to task graphs with
event-driven synchronization and an on-GPU scheduler~\cite{mpk2025}, using a
multi-level superoptimizer~\cite{wu2025mirage}. Event Tensor adds shape- and
data-dependent dynamism~\cite{eventtensor2026}, Ada-MK performs offline graph
search~\cite{adamk2026}, and Fleet adds a chiplet-level task
tier~\cite{fleet2026}. These compilers reuse one program across models without
per-model forging. In our measurements MPK utilization declines as model size
decreases (\S\ref{sec:obs:cell}), and the corresponding implementation uses
constants derived from model dimensions, including greatest-common-divisor
constraints and predicates~(\S\ref{sec:obs:constants}). Schedule-based
compilers, learned search, graph substitution, tile-level languages, autotuning
and sparse abstractions similarly search predefined program
spaces~\cite{ragankelley2013halide,chen2018tvm,zhu2022roller,shi2023welder,zheng2020ansor,jia2019taso,tillet2019triton,wang2024ladder,osama2023streamk,ansel2014opentuner,insum2026};
for a megakernel, task decomposition, counter granularity and buffer layout are
compilation decisions.

\paragraph{Agent harnesses for kernels.} Several harnesses generate or optimize
GPU kernels automatically, through multi-agent decomposition~\cite{astra2025},
hardware feedback~\cite{cudaforge2025}, data-flow invariants with counterexample
feedback~\cite{argus2026}, formal verification of generated
CUDA~\cite{proofwright2025} and in-place operator optimization for PyTorch
models~\cite{kernelforge2026}, alongside broader agent programming
capability~\cite{li2022alphacode,romera2024funsearch,jimenez2024swebench}. All
search and replace implementations within an existing model or operator
structure. AutoMegaKernel (AMK) is the closest to ours: an agent-driven loop
generates a persistent cooperative kernel for Llama-family models, with a static
validator for deadlocking and racing schedules~\cite{amk2026}. AMK reports
4--8\% of the weight-bandwidth ceiling up to 1.1B parameters and trails cuBLAS
on datacenter GPUs, whereas \sys{} reports 50.5--85.9\% MBU from 0.6B to 13B
inside a serving engine. \rev{Its validator addresses schedule safety, while our
oracle also checks numerical outputs, arithmetic precision and memory-traffic
accounting;} a race-free schedule alone does not establish correctness of the
rotary embedding, as five kernels in our experiments show.

%% file: sec/08-conclusion.tex
\section{Conclusion}
\label{sec:concl}

We presented \sys{}, a framework that employs coding agents to generate model-specific megakernels for autoregressive decoding. \sys{} consists of a universal knowledge base containing ten progressive milestones and an independent mid-state test oracle. The milestones state what structure to build, and the oracle checks the per-layer mid-states the kernel produces as well as its final output, making the megakernel checkable and trustworthy. Across 14 decoding megakernel generation tasks on models of 0.6B--13B parameters, the generated megakernels reach 50.5--85.9\% model bandwidth utilization, a geometric mean of 1.21$\times$ over tuned SGLang and 1.54$\times$ over a megakernel compiler, and 1.11$\times$ decoding speedup inside SGLang at comparable accuracy. These results demonstrate the practical value of coupling reusable structural guidance with independent validation and hardware feedback, paving a path toward reducing reliance on model-specific manual kernel development while continually improving inference efficiency.

%% file: sec/abl-body.tex
\label{sec:eval:abl:milestone}

\S\ref{sec:design:ladder} requires structure and measured performance together,
and this experiment tests whether the structure earns its place. Each ablation
arm removes one of its three steps and keeps the MBU floor that step carries,
both in the gate and in the rules file the agent reads: the per-SM instruction
stream at \rung{M6}, the dependency counters at \rung{M7} and the shared-memory
buffer pool at \rung{M8}. Every arm
forges Qwen3-0.6B at $b1{\times}128$ from the same seeded scaffold, with the
same agent, prompt and thresholds, so the ladder is the only thing that
differs. The configuration's MBU ceiling is 39.87\%, and the retained
thresholds are 28\%, 35\% and 38\%. Absolute MBU does not travel between
machines here, so every arm runs beside a control that shares its host, its
cards and its measurement window, under a fixed budget of 50 rounds.

\begin{figure}[!ht]
\centering
\includegraphics[width=\columnwidth]{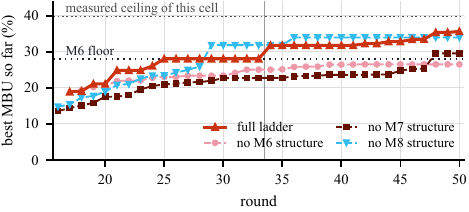}
\caption{Milestone ablation, best MBU so far against round. Each arm removes one
step of \S\ref{sec:design:ladder} and keeps the MBU floor it carries.}
\label{fig:abl}
\end{figure}

Figure~\ref{fig:abl} draws the four arms of the campaign. The control
clears the 28\% floor at round~25 and finishes at 35.69\%, just under 90\% of
what this configuration can reach. All three ablated arms finish below the floor
they kept. The arm without the instruction stream improves slowly throughout and
ends at 26.54\%. The arm without the counters gains most of its ground in the
last five rounds and still ends at 29.60\%, short of 35\%. The arm without the
buffer pool reaches 34.00\%, short of 38\%.

The instruction-stream requirement at \rung{M6} does real work on this
configuration. The control clears the 28\% floor while the arm that lost the
stream stops 1.5 points short of it on the same budget, and the two campaigns
differ in nothing else. The counter requirement at \rung{M7} points the same
way, though its arm was still climbing when the budget ran out, which leaves
room for it to be slower rather than blocked. \rung{M8} sets
the ceiling at the other end of the range, where a batched cell with a long
context makes the staged working set large enough to bound the step, and that is
where we would take the experiment next. The evidence here is one campaign on
one model at one shape, so it reports the direction of each requirement rather
than its size.

%% file: sec/supp-body.tex
\label{sec:eval:supp:agent}

Every campaign above is driven by the same coding agent. To check that \sys{}
does not depend on it, we re-forge Qwen3-0.6B at $b1{\times}128$ with
\code{codex exec} on GPT-5.6-sol, leaving the seeded scaffold, the rungs, the
thresholds and the prompt unchanged.

\input{sec/tab-agent}

Table~\ref{tab:agent} puts that run beside the strongest of the four
default-harness runs of the same cell. The campaign climbs the ladder unaided and
produces a working megakernel at round~44 of 84: one block per SM, a
14{,}916-instruction interpreter, and a decode step of 1009.6\us{} end to end, or
35.7\% MBU, 89.5\% of the cell's ceiling and $1.73\times$ SGLang and
$1.50\times$ MPK on the same card. The default-harness runs reach 26.4\% to
35.0\% MBU for 194 to 378\,Mtok, so the ladder and its oracle transfer to a
different agent at comparable token cost.

%% file: sec/tab-agent.tex
\begin{table}[t]
\centering
\footnotesize
\setlength{\tabcolsep}{4pt}
\caption{Qwen3-0.6B at $b1{\times}128$, ceiling 39.87\% MBU, forged by
two coding agents.}
\label{tab:agent}
\begin{tabular}{@{}lrrrr@{}}
\toprule
Agent & rounds & MBU (\%) & Mtok & cost \\
\midrule
Claude Code, Claude Opus 5 1M & 55 & 35.0 & 378 & \$362 \\
\code{codex exec}, GPT-5.6-sol & 84 & 35.7 & 279 & n/a \\
\bottomrule
\end{tabular}
\end{table}

%% file: forgemegakernel.bbl

\begin{thebibliography}{58}


\ifx \showCODEN    \undefined \def \showCODEN     #1{\unskip}     \fi
\ifx \showISBNx    \undefined \def \showISBNx     #1{\unskip}     \fi
\ifx \showISBNxiii \undefined \def \showISBNxiii  #1{\unskip}     \fi
\ifx \showISSN     \undefined \def \showISSN      #1{\unskip}     \fi
\ifx \showLCCN     \undefined \def \showLCCN      #1{\unskip}     \fi
\ifx \shownote     \undefined \def \shownote      #1{#1}          \fi
\ifx \showarticletitle \undefined \def \showarticletitle #1{#1}   \fi
\ifx \showURL      \undefined \def \showURL       {\relax}        \fi
\providecommand\bibfield[2]{#2}
\providecommand\bibinfo[2]{#2}
\providecommand\natexlab[1]{#1}
\providecommand\showeprint[2][]{arXiv:#2}

\bibitem[Gupta et~al\mbox{.}()]%
        {gupta2012persistent}
\bibfield{author}{\bibinfo{person}{Kshitij Gupta}, \bibinfo{person}{Jeff~A.
  Stuart}, {and} \bibinfo{person}{John~D. Owens}.}
  \bibinfo{year}{2012}\natexlab{}.
\newblock \showarticletitle{A Study of Persistent Threads Style {GPU}
  Programming for {GPGPU} Workloads}. In \bibinfo{booktitle}{\emph{Innovative
  Parallel Computing (InPar)}}. \bibinfo{address}{San Jose, CA, USA},
  \bibinfo{pages}{1--14}.
\newblock


\bibitem[Laine et~al\mbox{.}()]%
        {laine2013megakernels}
\bibfield{author}{\bibinfo{person}{Samuli Laine}, \bibinfo{person}{Tero
  Karras}, {and} \bibinfo{person}{Timo Aila}.} \bibinfo{year}{2013}\natexlab{}.
\newblock \showarticletitle{Megakernels Considered Harmful: Wavefront Path
  Tracing on {GPUs}}. In \bibinfo{booktitle}{\emph{Proceedings of the 5th
  High-Performance Graphics Conference (HPG)}}. \bibinfo{address}{Anaheim, CA,
  USA}, \bibinfo{pages}{137--143}.
\newblock


\bibitem[Yu et~al\mbox{.}()]%
        {yu2022orca}
\bibfield{author}{\bibinfo{person}{Gyeong-In Yu}, \bibinfo{person}{Joo~Seong
  Jeong}, \bibinfo{person}{Geon-Woo Kim}, \bibinfo{person}{Soojeong Kim}, {and}
  \bibinfo{person}{Byung-Gon Chun}.} \bibinfo{year}{2022}\natexlab{}.
\newblock \showarticletitle{{Orca}: A Distributed Serving System for
  Transformer-Based Generative Models}. In \bibinfo{booktitle}{\emph{16th
  USENIX Symposium on Operating Systems Design and Implementation (OSDI)}}.
  \bibinfo{address}{Carlsbad, CA, USA}, \bibinfo{pages}{521--538}.
\newblock


\bibitem[Agrawal et~al\mbox{.}()]%
        {agrawal2024sarathi}
\bibfield{author}{\bibinfo{person}{Amey Agrawal}, \bibinfo{person}{Nitin
  Kedia}, \bibinfo{person}{Ashish Panwar}, \bibinfo{person}{Jayashree Mohan},
  \bibinfo{person}{Nipun Kwatra}, \bibinfo{person}{Bhargav~S. Gulavani},
  \bibinfo{person}{Alexey Tumanov}, {and} \bibinfo{person}{Ramachandran
  Ramjee}.} \bibinfo{year}{2024}\natexlab{}.
\newblock \showarticletitle{Taming Throughput-Latency Tradeoff in {LLM}
  Inference with {Sarathi-Serve}}. In \bibinfo{booktitle}{\emph{18th USENIX
  Symposium on Operating Systems Design and Implementation (OSDI)}}.
  \bibinfo{address}{Santa Clara, CA, USA}, \bibinfo{pages}{117--134}.
\newblock


\bibitem[Kwon et~al\mbox{.}()]%
        {kwon2023pagedattention}
\bibfield{author}{\bibinfo{person}{Woosuk Kwon}, \bibinfo{person}{Zhuohan Li},
  \bibinfo{person}{Siyuan Zhuang}, \bibinfo{person}{Ying Sheng},
  \bibinfo{person}{Lianmin Zheng}, \bibinfo{person}{Cody~Hao Yu},
  \bibinfo{person}{Joseph~E. Gonzalez}, \bibinfo{person}{Hao Zhang}, {and}
  \bibinfo{person}{Ion Stoica}.} \bibinfo{year}{2023}\natexlab{}.
\newblock \showarticletitle{Efficient Memory Management for Large Language
  Model Serving with {PagedAttention}}. In
  \bibinfo{booktitle}{\emph{Proceedings of the 29th Symposium on Operating
  Systems Principles (SOSP)}}. \bibinfo{publisher}{ACM},
  \bibinfo{address}{Koblenz, Germany}, \bibinfo{pages}{611--626}.
\newblock


\bibitem[Zheng et~al\mbox{.}()]%
        {zheng2024sglang}
\bibfield{author}{\bibinfo{person}{Lianmin Zheng}, \bibinfo{person}{Liangsheng
  Yin}, \bibinfo{person}{Zhiqiang Xie}, \bibinfo{person}{Chuyue~Livia Sun},
  \bibinfo{person}{Jeff Huang}, \bibinfo{person}{Cody~Hao Yu},
  \bibinfo{person}{Shiyi Cao}, \bibinfo{person}{Christos Kozyrakis},
  \bibinfo{person}{Ion Stoica}, \bibinfo{person}{Joseph~E. Gonzalez},
  \bibinfo{person}{Clark Barrett}, {and} \bibinfo{person}{Ying Sheng}.}
  \bibinfo{year}{2024}\natexlab{}.
\newblock \showarticletitle{{SGLang}: Efficient Execution of Structured
  Language Model Programs}. In \bibinfo{booktitle}{\emph{Advances in Neural
  Information Processing Systems (NeurIPS)}}. \bibinfo{address}{Vancouver,
  Canada}.
\newblock


\bibitem[{NVIDIA Corporation}(a)]%
        {nvidia2025cublas}
\bibfield{author}{\bibinfo{person}{{NVIDIA Corporation}}.}
  \bibinfo{year}{2025}\natexlab{a}.
\newblock \bibinfo{title}{{cuBLAS} Library User Guide, Version 12.8}.
\newblock \bibinfo{howpublished}{\url{https://docs.nvidia.com/cuda/cublas/}}.
\newblock


\bibitem[{NVIDIA Corporation}(b)]%
        {nvidia2025cutlass}
\bibfield{author}{\bibinfo{person}{{NVIDIA Corporation}}.}
  \bibinfo{year}{2025}\natexlab{b}.
\newblock \bibinfo{title}{{CUTLASS}: {CUDA} Templates for Linear Algebra
  Subroutines}.
\newblock \bibinfo{howpublished}{\url{https://github.com/NVIDIA/cutlass}}.
\newblock


\bibitem[Tillet et~al\mbox{.}()]%
        {tillet2019triton}
\bibfield{author}{\bibinfo{person}{Philippe Tillet},
  \bibinfo{person}{Hsiang-Tsung Kung}, {and} \bibinfo{person}{David Cox}.}
  \bibinfo{year}{2019}\natexlab{}.
\newblock \showarticletitle{{Triton}: An Intermediate Language and Compiler for
  Tiled Neural Network Computations}. In \bibinfo{booktitle}{\emph{Proceedings
  of the 3rd ACM SIGPLAN International Workshop on Machine Learning and
  Programming Languages (MAPL)}}. \bibinfo{address}{Phoenix, AZ, USA},
  \bibinfo{pages}{10--19}.
\newblock


\bibitem[Ye et~al\mbox{.}()]%
        {ye2025flashinfer}
\bibfield{author}{\bibinfo{person}{Zihao Ye}, \bibinfo{person}{Lequn Chen},
  \bibinfo{person}{Ruihang Lai}, \bibinfo{person}{Wuwei Lin},
  \bibinfo{person}{Yineng Zhang}, \bibinfo{person}{Stephanie Wang},
  \bibinfo{person}{Tianqi Chen}, \bibinfo{person}{Baris Kasikci},
  \bibinfo{person}{Vinod Grover}, \bibinfo{person}{Arvind Krishnamurthy}, {and}
  \bibinfo{person}{Luis Ceze}.} \bibinfo{year}{2025}\natexlab{}.
\newblock \showarticletitle{{FlashInfer}: Efficient and Customizable Attention
  Engine for {LLM} Inference Serving}. In \bibinfo{booktitle}{\emph{Proceedings
  of the Eighth Annual Conference on Machine Learning and Systems (MLSys)}}.
  \bibinfo{address}{Santa Clara, CA, USA}.
\newblock


\bibitem[Spector et~al\mbox{.}()]%
        {spector2025nobubbles}
\bibfield{author}{\bibinfo{person}{Benjamin~F. Spector},
  \bibinfo{person}{Jordan Juravsky}, \bibinfo{person}{Stuart Sul},
  \bibinfo{person}{Owen Dugan}, \bibinfo{person}{Dylan Lim},
  \bibinfo{person}{Daniel~Y. Fu}, \bibinfo{person}{Simran Arora}, {and}
  \bibinfo{person}{Christopher R{\'e}}.} \bibinfo{year}{2025}\natexlab{}.
\newblock \bibinfo{title}{Look {Ma}, No Bubbles! Designing a Low-Latency
  Megakernel for {Llama-1B}}.
\newblock \bibinfo{howpublished}{Hazy Research blog,
  \url{https://hazyresearch.stanford.edu/blog/2025-05-27-no-bubbles}}.
\newblock


\bibitem[Nrusimha et~al\mbox{.}()]%
        {nrusimha2025flashformer}
\bibfield{author}{\bibinfo{person}{Aniruddha Nrusimha},
  \bibinfo{person}{William Brandon}, \bibinfo{person}{Mayank Mishra},
  \bibinfo{person}{Yikang Shen}, \bibinfo{person}{Rameswar Panda},
  \bibinfo{person}{Jonathan Ragan-Kelley}, {and} \bibinfo{person}{Yoon Kim}.}
  \bibinfo{year}{2025}\natexlab{}.
\newblock \bibinfo{title}{{FlashFormer}: Whole-Model Kernels for Efficient
  Low-Batch Inference}.
\newblock \bibinfo{howpublished}{arXiv:2505.22758}.
\newblock


\bibitem[Dao et~al\mbox{.}()]%
        {dao2022flashattention}
\bibfield{author}{\bibinfo{person}{Tri Dao}, \bibinfo{person}{Daniel~Y. Fu},
  \bibinfo{person}{Stefano Ermon}, \bibinfo{person}{Atri Rudra}, {and}
  \bibinfo{person}{Christopher R{\'e}}.} \bibinfo{year}{2022}\natexlab{}.
\newblock \showarticletitle{{FlashAttention}: Fast and Memory-Efficient Exact
  Attention with {IO}-Awareness}. In \bibinfo{booktitle}{\emph{Advances in
  Neural Information Processing Systems (NeurIPS)}}. \bibinfo{address}{New
  Orleans, LA, USA}.
\newblock


\bibitem[Dao()]%
        {dao2024flashattention2}
\bibfield{author}{\bibinfo{person}{Tri Dao}.} \bibinfo{year}{2024}\natexlab{}.
\newblock \showarticletitle{{FlashAttention-2}: Faster Attention with Better
  Parallelism and Work Partitioning}. In \bibinfo{booktitle}{\emph{The Twelfth
  International Conference on Learning Representations (ICLR)}}.
  \bibinfo{address}{Vienna, Austria}.
\newblock


\bibitem[Spector et~al\mbox{.}()]%
        {spector2025thunderkittens}
\bibfield{author}{\bibinfo{person}{Benjamin~F. Spector},
  \bibinfo{person}{Simran Arora}, \bibinfo{person}{Aaryan Singhal},
  \bibinfo{person}{Arjun Parthasarathy}, \bibinfo{person}{Daniel~Y. Fu}, {and}
  \bibinfo{person}{Christopher R{\'e}}.} \bibinfo{year}{2025}\natexlab{}.
\newblock \showarticletitle{{ThunderKittens}: Simple, Fast, and Adorable {AI}
  Kernels}. In \bibinfo{booktitle}{\emph{The Thirteenth International
  Conference on Learning Representations (ICLR)}}.
  \bibinfo{address}{Singapore}.
\newblock


\bibitem[Cheng et~al\mbox{.}()]%
        {mpk2025}
\bibfield{author}{\bibinfo{person}{Xinhao Cheng}, \bibinfo{person}{Zhihao
  Zhang}, \bibinfo{person}{Yu Zhou}, \bibinfo{person}{Jianan Ji},
  \bibinfo{person}{Jinchen Jiang}, \bibinfo{person}{Zepeng Zhao},
  \bibinfo{person}{Ziruo Xiao}, \bibinfo{person}{Zihao Ye},
  \bibinfo{person}{Yingyi Huang}, \bibinfo{person}{Ruihang Lai},
  \bibinfo{person}{Hongyi Jin}, \bibinfo{person}{Bohan Hou},
  \bibinfo{person}{Mengdi Wu}, \bibinfo{person}{Yixin Dong},
  \bibinfo{person}{Anthony Yip}, \bibinfo{person}{Songting Wang},
  \bibinfo{person}{Wenqin Yang}, \bibinfo{person}{Xupeng Miao},
  \bibinfo{person}{Tianqi Chen}, {and} \bibinfo{person}{Zhihao Jia}.}
  \bibinfo{year}{2025}\natexlab{}.
\newblock \bibinfo{title}{{MPK}: A Compiler and Runtime for Mega-Kernelizing
  Tensor Programs}.
\newblock \bibinfo{howpublished}{arXiv:2512.22219}.
\newblock


\bibitem[Jin et~al\mbox{.}()]%
        {eventtensor2026}
\bibfield{author}{\bibinfo{person}{Hongyi Jin}, \bibinfo{person}{Bohan Hou},
  \bibinfo{person}{Guanjie Wang}, \bibinfo{person}{Ruihang Lai},
  \bibinfo{person}{Jinqi Chen}, \bibinfo{person}{Zihao Ye},
  \bibinfo{person}{Yaxing Cai}, \bibinfo{person}{Yixin Dong},
  \bibinfo{person}{Xinhao Cheng}, \bibinfo{person}{Zhihao Zhang},
  \bibinfo{person}{Yilong Zhao}, \bibinfo{person}{Yingyi Huang},
  \bibinfo{person}{Lijie Yang}, \bibinfo{person}{Jinchen Jiang},
  \bibinfo{person}{Gabriele Oliaro}, \bibinfo{person}{Jianan Ji},
  \bibinfo{person}{Xupeng Miao}, \bibinfo{person}{Vinod Grover},
  \bibinfo{person}{Todd~C. Mowry}, \bibinfo{person}{Zhihao Jia}, {and}
  \bibinfo{person}{Tianqi Chen}.} \bibinfo{year}{2026}\natexlab{}.
\newblock \bibinfo{title}{Event Tensor: A Unified Abstraction for Compiling
  Dynamic Megakernels}.
\newblock \bibinfo{howpublished}{arXiv:2604.13327}.
\newblock


\bibitem[Dong et~al\mbox{.}()]%
        {adamk2026}
\bibfield{author}{\bibinfo{person}{Wenxin Dong}, \bibinfo{person}{Mingqing Hu},
  \bibinfo{person}{Guanghui Yu}, \bibinfo{person}{Qiang Fu},
  \bibinfo{person}{Peng Xu}, \bibinfo{person}{Hui Xu}, \bibinfo{person}{Yue
  Xing}, \bibinfo{person}{Xuewu Jiao}, \bibinfo{person}{Shuanglong Li}, {and}
  \bibinfo{person}{Lin Liu}.} \bibinfo{year}{2026}\natexlab{}.
\newblock \bibinfo{title}{{Ada-MK}: Adaptive MegaKernel Optimization via
  Automated {DAG}-based Search for {LLM} Inference}.
\newblock \bibinfo{howpublished}{arXiv:2605.11581}.
\newblock


\bibitem[Ouyang et~al\mbox{.}()]%
        {ouyang2025kernelbench}
\bibfield{author}{\bibinfo{person}{Anne Ouyang}, \bibinfo{person}{Simon Guo},
  \bibinfo{person}{Simran Arora}, \bibinfo{person}{Alex~L. Zhang},
  \bibinfo{person}{William Hu}, \bibinfo{person}{Christopher R{\'e}}, {and}
  \bibinfo{person}{Azalia Mirhoseini}.} \bibinfo{year}{2025}\natexlab{}.
\newblock \bibinfo{title}{{KernelBench}: Can {LLMs} Write Efficient {GPU}
  Kernels?}
\newblock \bibinfo{howpublished}{arXiv:2502.10517}.
\newblock


\bibitem[Wei et~al\mbox{.}()]%
        {astra2025}
\bibfield{author}{\bibinfo{person}{Anjiang Wei}, \bibinfo{person}{Tianran Sun},
  \bibinfo{person}{Yogesh Seenichamy}, \bibinfo{person}{Hang Song},
  \bibinfo{person}{Anne Ouyang}, \bibinfo{person}{Azalia Mirhoseini},
  \bibinfo{person}{Ke Wang}, {and} \bibinfo{person}{Alex Aiken}.}
  \bibinfo{year}{2025}\natexlab{}.
\newblock \bibinfo{title}{{Astra}: A Multi-Agent System for {GPU} Kernel
  Performance Optimization}.
\newblock \bibinfo{howpublished}{arXiv:2509.07506}.
\newblock


\bibitem[Zhang et~al\mbox{.}()]%
        {cudaforge2025}
\bibfield{author}{\bibinfo{person}{Zijian Zhang}, \bibinfo{person}{Rong Wang},
  \bibinfo{person}{Shiyang Li}, \bibinfo{person}{Yuebo Luo},
  \bibinfo{person}{Mingyi Hong}, {and} \bibinfo{person}{Caiwen Ding}.}
  \bibinfo{year}{2025}\natexlab{}.
\newblock \bibinfo{title}{{CudaForge}: An Agent Framework with Hardware
  Feedback for {CUDA} Kernel Optimization}.
\newblock \bibinfo{howpublished}{arXiv:2511.01884}.
\newblock


\bibitem[Brodsky et~al\mbox{.}()]%
        {kernelforge2026}
\bibfield{author}{\bibinfo{person}{Joshua Brodsky}, \bibinfo{person}{Dhravid
  Kumar}, \bibinfo{person}{Savini Kashmira}, \bibinfo{person}{Jayanaka
  Danatanarayana}, \bibinfo{person}{Jason Mars}, \bibinfo{person}{Krisztian
  Flautner}, {and} \bibinfo{person}{Lingjia Tang}.}
  \bibinfo{year}{2026}\natexlab{}.
\newblock \bibinfo{title}{Kernel Forge: An Agent Harness for {LLM}-based
  Generation and Optimization of {CUDA} Kernels}.
\newblock \bibinfo{howpublished}{arXiv:2607.24762}.
\newblock


\bibitem[Kamath et~al\mbox{.}()]%
        {kamath2025podattention}
\bibfield{author}{\bibinfo{person}{Aditya~K. Kamath}, \bibinfo{person}{Ramya
  Prabhu}, \bibinfo{person}{Jayashree Mohan}, \bibinfo{person}{Simon Peter},
  \bibinfo{person}{Ramachandran Ramjee}, {and} \bibinfo{person}{Ashish
  Panwar}.} \bibinfo{year}{2025}\natexlab{}.
\newblock \showarticletitle{{POD-Attention}: Unlocking Full Prefill-Decode
  Overlap for Faster {LLM} Inference}. In \bibinfo{booktitle}{\emph{Proceedings
  of the 30th ACM International Conference on Architectural Support for
  Programming Languages and Operating Systems (ASPLOS)}}.
  \bibinfo{address}{Rotterdam, Netherlands}, \bibinfo{pages}{897--912}.
\newblock


\bibitem[Vaswani et~al\mbox{.}()]%
        {vaswani2017attention}
\bibfield{author}{\bibinfo{person}{Ashish Vaswani}, \bibinfo{person}{Noam
  Shazeer}, \bibinfo{person}{Niki Parmar}, \bibinfo{person}{Jakob Uszkoreit},
  \bibinfo{person}{Llion Jones}, \bibinfo{person}{Aidan~N. Gomez},
  \bibinfo{person}{Lukasz Kaiser}, {and} \bibinfo{person}{Illia Polosukhin}.}
  \bibinfo{year}{2017}\natexlab{}.
\newblock \showarticletitle{Attention Is All You Need}. In
  \bibinfo{booktitle}{\emph{Advances in Neural Information Processing Systems
  (NeurIPS)}}.
\newblock


\bibitem[Su et~al\mbox{.}()]%
        {su2024rope}
\bibfield{author}{\bibinfo{person}{Jianlin Su}, \bibinfo{person}{Murtadha
  Ahmed}, \bibinfo{person}{Yu Lu}, \bibinfo{person}{Shengfeng Pan},
  \bibinfo{person}{Wen Bo}, {and} \bibinfo{person}{Yunfeng Liu}.}
  \bibinfo{year}{2024}\natexlab{}.
\newblock \showarticletitle{{RoFormer}: Enhanced Transformer with Rotary
  Position Embedding}.
\newblock \bibinfo{journal}{\emph{Neurocomputing}}  \bibinfo{volume}{568}
  (\bibinfo{year}{2024}), \bibinfo{pages}{127063}.
\newblock


\bibitem[Shazeer()]%
        {shazeer2019mqa}
\bibfield{author}{\bibinfo{person}{Noam Shazeer}.}
  \bibinfo{year}{2019}\natexlab{}.
\newblock \showarticletitle{Fast Transformer Decoding: One Write-Head is All
  You Need}.
\newblock \bibinfo{journal}{\emph{arXiv:1911.02150}} (\bibinfo{year}{2019}).
\newblock


\bibitem[Ainslie et~al\mbox{.}()]%
        {ainslie2023gqa}
\bibfield{author}{\bibinfo{person}{Joshua Ainslie}, \bibinfo{person}{James
  Lee-Thorp}, \bibinfo{person}{Michiel de Jong}, \bibinfo{person}{Yury
  Zemlyanskiy}, \bibinfo{person}{Federico Lebr{\'o}n}, {and}
  \bibinfo{person}{Sumit Sanghai}.} \bibinfo{year}{2023}\natexlab{}.
\newblock \showarticletitle{{GQA}: Training Generalized Multi-Query Transformer
  Models from Multi-Head Checkpoints}. In \bibinfo{booktitle}{\emph{Proceedings
  of the 2023 Conference on Empirical Methods in Natural Language Processing
  (EMNLP)}}. \bibinfo{address}{Singapore}, \bibinfo{pages}{4895--4901}.
\newblock


\bibitem[Agarwal et~al\mbox{.}()]%
        {databricks2023mbu}
\bibfield{author}{\bibinfo{person}{Megha Agarwal}, \bibinfo{person}{Asfandyar
  Qureshi}, \bibinfo{person}{Nikhil Sardana}, \bibinfo{person}{Linden Li},
  \bibinfo{person}{Julian Quevedo}, {and} \bibinfo{person}{Daya Khudia}.}
  \bibinfo{year}{2023}\natexlab{}.
\newblock \bibinfo{title}{{LLM} Inference Performance Engineering: Best
  Practices}.
\newblock \bibinfo{howpublished}{Databricks Engineering Blog,
  \url{https://www.databricks.com/blog/llm-inference-performance-engineering-best-practices}}.
\newblock


\bibitem[{NVIDIA Corporation}()]%
        {nvidia2024cudagraphs}
\bibfield{author}{\bibinfo{person}{{NVIDIA Corporation}}.}
  \bibinfo{year}{2025}\natexlab{}.
\newblock \bibinfo{title}{{CUDA} {C}++ Programming Guide, Version 12.8}.
\newblock
  \bibinfo{howpublished}{\url{https://docs.nvidia.com/cuda/cuda-c-programming-guide/}}.
\newblock


\bibitem[Williams et~al\mbox{.}()]%
        {williams2009roofline}
\bibfield{author}{\bibinfo{person}{Samuel Williams}, \bibinfo{person}{Andrew
  Waterman}, {and} \bibinfo{person}{David Patterson}.}
  \bibinfo{year}{2009}\natexlab{}.
\newblock \showarticletitle{Roofline: An Insightful Visual Performance Model
  for Multicore Architectures}.
\newblock \bibinfo{journal}{\emph{Commun. ACM}} \bibinfo{volume}{52},
  \bibinfo{number}{4} (\bibinfo{year}{2009}), \bibinfo{pages}{65--76}.
\newblock


\bibitem[{Anthropic}()]%
        {claudecode2025}
\bibfield{author}{\bibinfo{person}{{Anthropic}}.}
  \bibinfo{year}{2025}\natexlab{}.
\newblock \bibinfo{title}{{Claude Code}: An Agentic Command-Line Tool for
  Software Engineering}.
\newblock \bibinfo{howpublished}{\url{https://www.anthropic.com/claude-code}}.
\newblock


\bibitem[Gerganov and {llama.cpp contributors}()]%
        {llamacpp}
\bibfield{author}{\bibinfo{person}{Georgi Gerganov} {and}
  \bibinfo{person}{{llama.cpp contributors}}.} \bibinfo{year}{2026}\natexlab{}.
\newblock \bibinfo{title}{{llama.cpp}: {LLM} inference in {C/C++}}.
\newblock \bibinfo{howpublished}{\url{https://github.com/ggml-org/llama.cpp}}.
\newblock


\bibitem[Grattafiori et~al\mbox{.}()]%
        {grattafiori2024llama3}
\bibfield{author}{\bibinfo{person}{Aaron Grattafiori},
  \bibinfo{person}{Abhimanyu Dubey}, \bibinfo{person}{Abhinav Jauhri},
  \bibinfo{person}{Abhinav Pandey}, \bibinfo{person}{Abhishek Kadian},
  \bibinfo{person}{Ahmad Al-Dahle}, \bibinfo{person}{Aiesha Letman},
  \bibinfo{person}{Akhil Mathur}, \bibinfo{person}{Alan Schelten}, {and}
  \bibinfo{person}{Alex Vaughan}.} \bibinfo{year}{2024}\natexlab{}.
\newblock \bibinfo{title}{The {Llama 3} Herd of Models}.
\newblock \bibinfo{howpublished}{arXiv:2407.21783}.
\newblock


\bibitem[Touvron et~al\mbox{.}()]%
        {touvron2023llama2}
\bibfield{author}{\bibinfo{person}{Hugo Touvron}, \bibinfo{person}{Louis
  Martin}, \bibinfo{person}{Kevin Stone}, \bibinfo{person}{Peter Albert},
  \bibinfo{person}{Amjad Almahairi}, \bibinfo{person}{Yasmine Babaei},
  \bibinfo{person}{Nikolay Bashlykov}, \bibinfo{person}{Soumya Batra},
  \bibinfo{person}{Prajjwal Bhargava}, {and} \bibinfo{person}{Shruti Bhosale}.}
  \bibinfo{year}{2023}\natexlab{}.
\newblock \bibinfo{title}{{Llama 2}: Open Foundation and Fine-Tuned Chat
  Models}.
\newblock \bibinfo{howpublished}{arXiv:2307.09288}.
\newblock


\bibitem[Yang et~al\mbox{.}()]%
        {yang2025qwen3}
\bibfield{author}{\bibinfo{person}{An Yang}, \bibinfo{person}{Anfeng Li},
  \bibinfo{person}{Baosong Yang}, \bibinfo{person}{Beichen Zhang},
  \bibinfo{person}{Binyuan Hui}, \bibinfo{person}{Bo Zheng},
  \bibinfo{person}{Bowen Yu}, \bibinfo{person}{Chang Gao},
  \bibinfo{person}{Chengen Huang}, {and} \bibinfo{person}{Chenxu Lv}.}
  \bibinfo{year}{2025}\natexlab{}.
\newblock \bibinfo{title}{{Qwen3} Technical Report}.
\newblock \bibinfo{howpublished}{arXiv:2505.09388}.
\newblock


\bibitem[Hu et~al\mbox{.}()]%
        {hu2024minicpm}
\bibfield{author}{\bibinfo{person}{Shengding Hu}, \bibinfo{person}{Yuge Tu},
  \bibinfo{person}{Xu Han}, \bibinfo{person}{Chaoqun He},
  \bibinfo{person}{Ganqu Cui}, \bibinfo{person}{Xiang Long},
  \bibinfo{person}{Zhi Zheng}, \bibinfo{person}{Yewei Fang},
  \bibinfo{person}{Yuxiang Huang}, {and} \bibinfo{person}{Weilin Zhao}.}
  \bibinfo{year}{2024}\natexlab{}.
\newblock \bibinfo{title}{{MiniCPM}: Unveiling the Potential of Small Language
  Models with Scalable Training Strategies}.
\newblock \bibinfo{howpublished}{arXiv:2404.06395}.
\newblock


\bibitem[Cobbe et~al\mbox{.}()]%
        {cobbe2021gsm8k}
\bibfield{author}{\bibinfo{person}{Karl Cobbe}, \bibinfo{person}{Vineet
  Kosaraju}, \bibinfo{person}{Mohammad Bavarian}, \bibinfo{person}{Mark Chen},
  \bibinfo{person}{Heewoo Jun}, \bibinfo{person}{Lukasz Kaiser},
  \bibinfo{person}{Matthias Plappert}, \bibinfo{person}{Jerry Tworek},
  \bibinfo{person}{Jacob Hilton}, \bibinfo{person}{Reiichiro Nakano},
  \bibinfo{person}{Christopher Hesse}, {and} \bibinfo{person}{John Schulman}.}
  \bibinfo{year}{2021}\natexlab{}.
\newblock \showarticletitle{Training Verifiers to Solve Math Word Problems}.
\newblock \bibinfo{journal}{\emph{arXiv preprint arXiv:2110.14168}}
  (\bibinfo{year}{2021}).
\newblock


\bibitem[Juravsky et~al\mbox{.}()]%
        {juravsky2025wholegpu}
\bibfield{author}{\bibinfo{person}{Jordan Juravsky}, \bibinfo{person}{Stuart
  Sul}, \bibinfo{person}{Benjamin~F. Spector}, {and}
  \bibinfo{person}{Christopher R{\'e}}.} \bibinfo{year}{2025}\natexlab{}.
\newblock \bibinfo{title}{One Kernel for All Your {GPUs}}.
\newblock \bibinfo{howpublished}{Hazy Research blog,
  \url{https://hazyresearch.stanford.edu/blog/2025-09-22-pgl}}.
\newblock


\bibitem[{Kog AI}()]%
        {kog2026monokernel}
\bibfield{author}{\bibinfo{person}{{Kog AI}}.} \bibinfo{year}{2026}\natexlab{}.
\newblock \bibinfo{title}{Building a Single-Kernel, Latency-Optimized {LLM}
  Inference Engine on {AMD} {MI300X} {GPUs}}.
\newblock
  \bibinfo{howpublished}{\url{https://blog.kog.ai/building-a-single-kernel-latency-optimized-llm-inference-engine-on-amd-mi300x-gpus/}}.
\newblock


\bibitem[Sul and R{\'e}()]%
        {retireabstractions}
\bibfield{author}{\bibinfo{person}{Stuart Sul} {and}
  \bibinfo{person}{Christopher R{\'e}}.} \bibinfo{year}{2026}\natexlab{}.
\newblock \bibinfo{title}{Retire the Abstractions}.
\newblock \bibinfo{howpublished}{Hazy Research blog,
  \url{https://hazyresearch.stanford.edu/blog/2026-08-05-retire-the-abstractions}}.
\newblock


\bibitem[Wu et~al\mbox{.}()]%
        {wu2025mirage}
\bibfield{author}{\bibinfo{person}{Mengdi Wu}, \bibinfo{person}{Xinhao Cheng},
  \bibinfo{person}{Shengyu Liu}, \bibinfo{person}{Chunan Shi},
  \bibinfo{person}{Jianan Ji}, \bibinfo{person}{Man~Kit Ao},
  \bibinfo{person}{Praveen Velliengiri}, \bibinfo{person}{Xupeng Miao},
  \bibinfo{person}{Oded Padon}, {and} \bibinfo{person}{Zhihao Jia}.}
  \bibinfo{year}{2025}\natexlab{}.
\newblock \showarticletitle{{Mirage}: A Multi-Level Superoptimizer for Tensor
  Programs}. In \bibinfo{booktitle}{\emph{19th USENIX Symposium on Operating
  Systems Design and Implementation (OSDI)}}. \bibinfo{address}{Boston, MA,
  USA}.
\newblock


\bibitem[Chowdhary et~al\mbox{.}()]%
        {fleet2026}
\bibfield{author}{\bibinfo{person}{Sangeeta Chowdhary}, \bibinfo{person}{Ryan
  Swann}, \bibinfo{person}{Sean Siddens}, \bibinfo{person}{Muhammad Osama},
  \bibinfo{person}{Stephen Neuendorffer}, \bibinfo{person}{Alexandru Dutu},
  \bibinfo{person}{Karthik Sangaiah}, \bibinfo{person}{Sandeepa Bhuyan},
  \bibinfo{person}{Samuel Bayliss}, {and} \bibinfo{person}{Ganesh Dasika}.}
  \bibinfo{year}{2026}\natexlab{}.
\newblock \bibinfo{title}{{Fleet}: Hierarchical Task-based Abstraction for
  Megakernels on Multi-Die {GPUs}}.
\newblock \bibinfo{howpublished}{arXiv:2604.15379}.
\newblock


\bibitem[Ragan-Kelley et~al\mbox{.}()]%
        {ragankelley2013halide}
\bibfield{author}{\bibinfo{person}{Jonathan Ragan-Kelley},
  \bibinfo{person}{Connelly Barnes}, \bibinfo{person}{Andrew Adams},
  \bibinfo{person}{Sylvain Paris}, \bibinfo{person}{Fr{\'e}do Durand}, {and}
  \bibinfo{person}{Saman Amarasinghe}.} \bibinfo{year}{2013}\natexlab{}.
\newblock \showarticletitle{{Halide}: A Language and Compiler for Optimizing
  Parallelism, Locality, and Recomputation in Image Processing Pipelines}. In
  \bibinfo{booktitle}{\emph{Proceedings of the 34th ACM SIGPLAN Conference on
  Programming Language Design and Implementation (PLDI)}}.
  \bibinfo{address}{Seattle, WA, USA}, \bibinfo{pages}{519--530}.
\newblock


\bibitem[Chen et~al\mbox{.}()]%
        {chen2018tvm}
\bibfield{author}{\bibinfo{person}{Tianqi Chen}, \bibinfo{person}{Thierry
  Moreau}, \bibinfo{person}{Ziheng Jiang}, \bibinfo{person}{Lianmin Zheng},
  \bibinfo{person}{Eddie Yan}, \bibinfo{person}{Meghan Cowan},
  \bibinfo{person}{Haichen Shen}, \bibinfo{person}{Leyuan Wang},
  \bibinfo{person}{Yuwei Hu}, \bibinfo{person}{Luis Ceze},
  \bibinfo{person}{Carlos Guestrin}, {and} \bibinfo{person}{Arvind
  Krishnamurthy}.} \bibinfo{year}{2018}\natexlab{}.
\newblock \showarticletitle{{TVM}: An Automated End-to-End Optimizing Compiler
  for Deep Learning}. In \bibinfo{booktitle}{\emph{13th USENIX Symposium on
  Operating Systems Design and Implementation (OSDI)}}.
  \bibinfo{address}{Carlsbad, CA, USA}, \bibinfo{pages}{578--594}.
\newblock


\bibitem[Zhu et~al\mbox{.}()]%
        {zhu2022roller}
\bibfield{author}{\bibinfo{person}{Hongyu Zhu}, \bibinfo{person}{Ruofan Wu},
  \bibinfo{person}{Yijia Diao}, \bibinfo{person}{Shanbin Ke},
  \bibinfo{person}{Haoyu Li}, \bibinfo{person}{Chen Zhang},
  \bibinfo{person}{Jilong Xue}, \bibinfo{person}{Lingxiao Ma},
  \bibinfo{person}{Yuqing Xia}, \bibinfo{person}{Wei Cui}, \bibinfo{person}{Fan
  Yang}, \bibinfo{person}{Mao Yang}, \bibinfo{person}{Lidong Zhou},
  \bibinfo{person}{Asaf Cidon}, {and} \bibinfo{person}{Gennady Pekhimenko}.}
  \bibinfo{year}{2022}\natexlab{}.
\newblock \showarticletitle{{ROLLER}: Fast and Efficient Tensor Compilation for
  Deep Learning}. In \bibinfo{booktitle}{\emph{USENIX Symposium on Operating
  Systems Design and Implementation (OSDI)}}.
\newblock


\bibitem[Shi et~al\mbox{.}()]%
        {shi2023welder}
\bibfield{author}{\bibinfo{person}{Yining Shi}, \bibinfo{person}{Zhi Yang},
  \bibinfo{person}{Jilong Xue}, \bibinfo{person}{Lingxiao Ma},
  \bibinfo{person}{Yuqing Xia}, \bibinfo{person}{Ziming Miao},
  \bibinfo{person}{Yuxiao Guo}, \bibinfo{person}{Fan Yang}, {and}
  \bibinfo{person}{Lidong Zhou}.} \bibinfo{year}{2023}\natexlab{}.
\newblock \showarticletitle{Welder: Scheduling Deep Learning Memory Access via
  Tile-graph}. In \bibinfo{booktitle}{\emph{USENIX Symposium on Operating
  Systems Design and Implementation (OSDI)}}.
\newblock


\bibitem[Zheng et~al\mbox{.}()]%
        {zheng2020ansor}
\bibfield{author}{\bibinfo{person}{Lianmin Zheng}, \bibinfo{person}{Chengfan
  Jia}, \bibinfo{person}{Minmin Sun}, \bibinfo{person}{Zhao Wu},
  \bibinfo{person}{Cody~Hao Yu}, \bibinfo{person}{Ameer Haj-Ali},
  \bibinfo{person}{Yida Wang}, \bibinfo{person}{Jun Yang},
  \bibinfo{person}{Danyang Zhuo}, \bibinfo{person}{Koushik Sen},
  \bibinfo{person}{Joseph~E. Gonzalez}, {and} \bibinfo{person}{Ion Stoica}.}
  \bibinfo{year}{2020}\natexlab{}.
\newblock \showarticletitle{{Ansor}: Generating High-Performance Tensor
  Programs for Deep Learning}. In \bibinfo{booktitle}{\emph{14th USENIX
  Symposium on Operating Systems Design and Implementation (OSDI)}}.
  \bibinfo{address}{Banff, Canada}, \bibinfo{pages}{863--879}.
\newblock


\bibitem[Jia et~al\mbox{.}()]%
        {jia2019taso}
\bibfield{author}{\bibinfo{person}{Zhihao Jia}, \bibinfo{person}{Oded Padon},
  \bibinfo{person}{James Thomas}, \bibinfo{person}{Todd Warszawski},
  \bibinfo{person}{Matei Zaharia}, {and} \bibinfo{person}{Alex Aiken}.}
  \bibinfo{year}{2019}\natexlab{}.
\newblock \showarticletitle{{TASO}: Optimizing Deep Learning Computation with
  Automatic Generation of Graph Substitutions}. In
  \bibinfo{booktitle}{\emph{Proceedings of the 27th ACM Symposium on Operating
  Systems Principles (SOSP)}}. \bibinfo{address}{Huntsville, Canada},
  \bibinfo{pages}{47--62}.
\newblock


\bibitem[Wang et~al\mbox{.}()]%
        {wang2024ladder}
\bibfield{author}{\bibinfo{person}{Lei Wang}, \bibinfo{person}{Lingxiao Ma},
  \bibinfo{person}{Shijie Cao}, \bibinfo{person}{Quanlu Zhang},
  \bibinfo{person}{Jilong Xue}, \bibinfo{person}{Yining Shi},
  \bibinfo{person}{Ningxin Zheng}, \bibinfo{person}{Ziming Miao},
  \bibinfo{person}{Fan Yang}, \bibinfo{person}{Ting Cao},
  \bibinfo{person}{Yuqing Yang}, {and} \bibinfo{person}{Mao Yang}.}
  \bibinfo{year}{2024}\natexlab{}.
\newblock \showarticletitle{Ladder: Enabling Efficient Low-Precision Deep
  Learning Computing through Hardware-aware Tensor Transformation}. In
  \bibinfo{booktitle}{\emph{USENIX Symposium on Operating Systems Design and
  Implementation (OSDI)}}.
\newblock


\bibitem[Osama et~al\mbox{.}()]%
        {osama2023streamk}
\bibfield{author}{\bibinfo{person}{Muhammad Osama}, \bibinfo{person}{Duane
  Merrill}, \bibinfo{person}{Cris Cecka}, \bibinfo{person}{Michael Garland},
  {and} \bibinfo{person}{John~D. Owens}.} \bibinfo{year}{2023}\natexlab{}.
\newblock \showarticletitle{{Stream-K}: Work-Centric Parallel Decomposition for
  Dense Matrix-Matrix Multiplication on the {GPU}}. In
  \bibinfo{booktitle}{\emph{ACM SIGPLAN Symposium on Principles and Practice of
  Parallel Programming (PPoPP)}}.
\newblock


\bibitem[Ansel et~al\mbox{.}()]%
        {ansel2014opentuner}
\bibfield{author}{\bibinfo{person}{Jason Ansel}, \bibinfo{person}{Shoaib
  Kamil}, \bibinfo{person}{Kalyan Veeramachaneni}, \bibinfo{person}{Jonathan
  Ragan-Kelley}, \bibinfo{person}{Jeffrey Bosboom}, \bibinfo{person}{Una-May
  O'Reilly}, {and} \bibinfo{person}{Saman Amarasinghe}.}
  \bibinfo{year}{2014}\natexlab{}.
\newblock \showarticletitle{{OpenTuner}: An Extensible Framework for Program
  Autotuning}. In \bibinfo{booktitle}{\emph{Proceedings of the 23rd
  International Conference on Parallel Architectures and Compilation (PACT)}}.
  \bibinfo{address}{Edmonton, Canada}, \bibinfo{pages}{303--316}.
\newblock


\bibitem[Won et~al\mbox{.}()]%
        {insum2026}
\bibfield{author}{\bibinfo{person}{Jaeyeon Won}, \bibinfo{person}{Willow
  Ahrens}, \bibinfo{person}{Joel~S. Emer}, {and} \bibinfo{person}{Saman
  Amarasinghe}.} \bibinfo{year}{2026}\natexlab{}.
\newblock \showarticletitle{{Insum}: Sparse {GPU} Kernels Simplified and
  Optimized with Indirect Einsums}. In \bibinfo{booktitle}{\emph{Proceedings of
  the 31st ACM International Conference on Architectural Support for
  Programming Languages and Operating Systems (ASPLOS)}}.
  \bibinfo{address}{Pittsburgh, PA, USA}.
\newblock


\bibitem[Mai et~al\mbox{.}()]%
        {argus2026}
\bibfield{author}{\bibinfo{person}{Haohui Mai}, \bibinfo{person}{Xiaoyan Guo},
  \bibinfo{person}{Xiangyun Ding}, \bibinfo{person}{Christos Kozyrakis}, {and}
  \bibinfo{person}{Binhang Yuan}.} \bibinfo{year}{2026}\natexlab{}.
\newblock \bibinfo{title}{{ARGUS}: Agentic {GPU} Optimization Guided by
  Data-Flow Invariants}.
\newblock \bibinfo{howpublished}{arXiv:2604.18616}.
\newblock


\bibitem[Chatterjee et~al\mbox{.}()]%
        {proofwright2025}
\bibfield{author}{\bibinfo{person}{Bodhisatwa Chatterjee},
  \bibinfo{person}{Drew Zagieboylo}, \bibinfo{person}{Sana Damani},
  \bibinfo{person}{Siva Kumar~Sastry Hari}, {and} \bibinfo{person}{Christos
  Kozyrakis}.} \bibinfo{year}{2025}\natexlab{}.
\newblock \bibinfo{title}{{ProofWright}: Towards Agentic Formal Verification of
  {CUDA}}.
\newblock \bibinfo{howpublished}{arXiv:2511.12294}.
\newblock


\bibitem[Li et~al\mbox{.}()]%
        {li2022alphacode}
\bibfield{author}{\bibinfo{person}{Yujia Li}, \bibinfo{person}{David Choi},
  \bibinfo{person}{Junyoung Chung}, \bibinfo{person}{Nate Kushman},
  \bibinfo{person}{Julian Schrittwieser}, \bibinfo{person}{R{\'e}mi Leblond},
  \bibinfo{person}{Tom Eccles}, \bibinfo{person}{James Keeling},
  \bibinfo{person}{Felix Gimeno}, \bibinfo{person}{Agustin~Dal Lago},
  \bibinfo{person}{Thomas Hubert}, \bibinfo{person}{Peter Choy},
  \bibinfo{person}{Cyprien de Masson~d'Autume}, \bibinfo{person}{Igor
  Babuschkin}, \bibinfo{person}{Xinyun Chen}, \bibinfo{person}{Po-Sen Huang},
  \bibinfo{person}{Johannes Welbl}, \bibinfo{person}{Sven Gowal},
  \bibinfo{person}{Alexey Cherepanov}, \bibinfo{person}{James Molloy},
  \bibinfo{person}{Daniel~J. Mankowitz}, \bibinfo{person}{Esme~Sutherland
  Robson}, \bibinfo{person}{Pushmeet Kohli}, \bibinfo{person}{Nando de
  Freitas}, \bibinfo{person}{Koray Kavukcuoglu}, {and} \bibinfo{person}{Oriol
  Vinyals}.} \bibinfo{year}{2022}\natexlab{}.
\newblock \showarticletitle{Competition-Level Code Generation with
  {AlphaCode}}.
\newblock \bibinfo{journal}{\emph{Science}} \bibinfo{volume}{378},
  \bibinfo{number}{6624} (\bibinfo{year}{2022}), \bibinfo{pages}{1092--1097}.
\newblock


\bibitem[Romera-Paredes et~al\mbox{.}()]%
        {romera2024funsearch}
\bibfield{author}{\bibinfo{person}{Bernardino Romera-Paredes},
  \bibinfo{person}{Mohammadamin Barekatain}, \bibinfo{person}{Alexander
  Novikov}, \bibinfo{person}{Matej Balog}, \bibinfo{person}{M.~Pawan Kumar},
  \bibinfo{person}{Emilien Dupont}, \bibinfo{person}{Francisco J.~R. Ruiz},
  \bibinfo{person}{Jordan~S. Ellenberg}, \bibinfo{person}{Pengming Wang},
  \bibinfo{person}{Omar Fawzi}, \bibinfo{person}{Pushmeet Kohli}, {and}
  \bibinfo{person}{Alhussein Fawzi}.} \bibinfo{year}{2024}\natexlab{}.
\newblock \showarticletitle{Mathematical Discoveries from Program Search with
  Large Language Models}.
\newblock \bibinfo{journal}{\emph{Nature}}  \bibinfo{volume}{625}
  (\bibinfo{year}{2024}), \bibinfo{pages}{468--475}.
\newblock


\bibitem[Jimenez et~al\mbox{.}()]%
        {jimenez2024swebench}
\bibfield{author}{\bibinfo{person}{Carlos~E. Jimenez}, \bibinfo{person}{John
  Yang}, \bibinfo{person}{Alexander Wettig}, \bibinfo{person}{Shunyu Yao},
  \bibinfo{person}{Kexin Pei}, \bibinfo{person}{Ofir Press}, {and}
  \bibinfo{person}{Karthik Narasimhan}.} \bibinfo{year}{2024}\natexlab{}.
\newblock \showarticletitle{{SWE-bench}: Can Language Models Resolve Real-World
  {GitHub} Issues?}. In \bibinfo{booktitle}{\emph{The Twelfth International
  Conference on Learning Representations (ICLR)}}. \bibinfo{address}{Vienna,
  Austria}.
\newblock


\bibitem[Jaber and Jaber()]%
        {amk2026}
\bibfield{author}{\bibinfo{person}{Jaber Jaber} {and} \bibinfo{person}{Osama
  Jaber}.} \bibinfo{year}{2026}\natexlab{}.
\newblock \bibinfo{title}{{AutoMegaKernel}: A Statically-Checked Agent Harness
  for Self-Retargeting Megakernel Synthesis}.
\newblock \bibinfo{howpublished}{arXiv:2606.09682}.
\newblock


\end{thebibliography}
